\documentclass[%
 aip,
jap,
amsmath,amssymb,
preprint,%
]{revtex4-1}

\usepackage[english]{babel}

\usepackage{amssymb}
\usepackage{amsmath}
\usepackage{graphicx}
\usepackage{dcolumn}
\usepackage[colorlinks=true, allcolors=blue]{hyperref}
\usepackage{natbib}
\usepackage[utf8]{inputenc}
\usepackage[T1]{fontenc}
\usepackage{mathptmx}
\usepackage{etoolbox}

\begin{document}


\title{Machine learning interatomic potential for predicting the thermal properties of uranium nitride}

\author{Beihan Chen}
 \affiliation{Department of Nuclear Engineering, The Pennsylvania State University, University Park, PA 16802, USA}

\author{Zilong Hua}
\affiliation{Idaho National Laboratory, Idaho Falls, ID 83415, USA}

\author{Jennifer K. Watkins}
\affiliation{Idaho National Laboratory, Idaho Falls, ID 83415, USA}

\author{Linu Malakkal}
\affiliation{Idaho National Laboratory, Idaho Falls, ID 83415, USA}

\author{Marat Khafizov}
\affiliation{Department of Mechanical and Aerospace Engineering, The Ohio State University, Columbus, OH 43210, USA}

\author{David H. Hurley}
\affiliation{Idaho National Laboratory, Idaho Falls, ID 83415, USA}

\author{Miaomiao Jin}%
 \email{mmjin@psu.edu}
\affiliation{Department of Nuclear Engineering, The Pennsylvania State University, University Park, PA 16802, USA}%

\begin{abstract}

We present a combined computational and experimental investigation of the thermal properties of uranium nitride (UN), focusing on the development of a machine learning interatomic potential (MLIP) using the moment tensor potential (MTP) framework. The MLIP was trained on density functional theory (DFT) data and validated against various quantities including energies, forces, elastic constants, phonon dispersion, and defect formation energies, achieving excellent agreement with DFT calculations, prior experimental results and our thermal conductivity measurement. The potential was then employed in molecular dynamics (MD) simulations to predict key thermal properties such as melting point, thermal expansion, specific heat, and thermal conductivity. To further assess model accuracy, we fabricated a UN sample and performed new thermal conductivity measurements representative of single-crystal properties, which showed strong agreement with the MLIP predictions. This work confirms the reliability and predictive capability of the developed potential for determining the thermal properties of UN.

\end{abstract}


\maketitle

\section{Introduction}

In comparison to the traditional uranium oxide nuclear fuel, uranium nitride (UN) fuel offers several benefits, including higher fissile density, higher thermal conductivity \cite{Hayes1990tc}, good compatibility with the PUREX process \cite{Jones2023}, and good compatibility with most cladding materials,resulting in an extended fuel cycle time \cite{Youinou2014}. Due to these advantages, UN is considered a strong candidate for long-term accident-tolerant fuel in light water reactors and a promising option for Generation IV nuclear reactors \cite{Ekberg2018}.Therefore, it is essential to understand the thermal behavior of UN. Several properties of UN, such as phonon dispersion \cite{Jackman1986}, elastic constants \cite{Salleh1986,Hayes1990gp}, thermal expansion \cite{LIU2023154215}, specific heat capacity \cite{Takahashi1971}, and thermal conductivity \cite{Ross1988,Hayes1990tc}, have been investigated experimentally and summarized by Hayes et al. \cite{Hayes1990,Hayes1990gp,Hayes1990tc,Hayes1990cp} and more recently by Miller et al. \cite{Miller2024}. In addition to experimental measurements, atomic scale modeling such as density functional theory (DFT) and molecular dynamics (MD) plays a crucial role in interpreting and complementing existing experimental data of UN properties.

As one of the most accurate quantum mechanical calculation methods of materials modeling, DFT has been widely used to investigate physical properties of UN \cite{Gryaznov2012,Evarestov2008,Kocevski2022_spin,Mei2013,Szpunar2020}.However, it requires global self-consistent convergence of the electronic structure, whose computational cost is $O(N^3)$, thereby constraining the feasible system size to a range of $N$=10-100 atoms \cite{Bowler2012,Thompson2015}. Compared to DFT, MD simulations offer a scalable approach for evaluating thermal properties at finite temperatures and larger length scales, overcoming the common size effects expected in DFT calculations \cite{Bowler2012}. However, the predictive capability of MD simulations critically depends on the quality of the interatomic potential. Over the years, several empirical interatomic potentials have been developed for UN. Kurosaki et al. \cite{Kurosaki2000sp, Kurosaki2000} constructed a Busing–Ida-type potential augmented by a Morse function to study  properties such as thermal expansion, specific heat capacity and thermal conductivity. Chen et al.\cite{Chen2010} developed another Morse-type potential to calculate the structural and elastic properties of UN. However, Morse potentials lack the ability to represent directional bonding, limiting their accuracy for UN \cite{Mishin2005}.  To address this limitation, an angular-dependent potential (ADP) was developed by Kuksin et al. \cite{Kuksin2016} using force-matching technique to investigate the diffusion of point defects. This ADP was later modified by Tseplyaev and Starikov \cite{Tseplyaev2016} to improve the accuracy of UN property descriptions at high pressures. Additionally, a many-bodied embedded-atom (EAM) potential combined with a pairwise Buckingham term for UN was developed by Kocevski et al. \cite{Kocevski2022} to investigate the elastic properties, specific heat capacity and Xenon diffusion in UN. Most recently, AbdulHameed et al. \cite{AbdulHameed2024} assessed the performance of Tseplyaev and Starikov's ADP (hereafter referred to as the Tseplyaev potential) \cite{Tseplyaev2016} and Kocevski et al.'s EAM  (hereafter referred to as the Kocevski potential) \cite{Kocevski2022} by comparing derived quantities, including UN thermal expansion, melting point, elastic properties, specific heat capacity, phonon properties and defect formation energies. They showed that the Tseplyaev potential excels in capturing the energetic aspects of UN, whereas the Kocevski potential performs better in modeling the structural properties. Given the differences among available interatomic potentials, a reliable potential is still desired that can ideally capture all these properties for predicative modeling of UN.

In recent years,machine learned interatomic potentials (MLIPs) have emerged as an effective alternative for developing quantum-accurate interatomic models. Developing an MLIP requires generating a diverse dataset of atomic configurations with corresponding energies, forces, and stresses from DFT calculations, which is then used to train the potential using a specific machine learning framework. A variety of MLIP training approaches have been proposed, such as the neural-network potential (NNP) \cite{Behler2007,Behler2011}, the Gaussian approximation potential (GAP) \cite{Bartk2010, Bartk2013}, the spectral neighbor analysis potential (SNAP) \cite{Thompson2015,Li2018}, and the moment tensor potential (MTP) \cite{Shapeev2016}. Among these, the MTP framework has demonstrated an excellent trade-off between accuracy and computational efficiency \cite{Zuo_2020}. Moreover, it has been successfully applied to predict the thermal properties in metal nitrides \cite{Bock2024}. Building on these advantages, we developed an MTP for UN based on a DFT-generated dataset. While a recent ML potential \cite{alzate} has been developed using a hierarchically interacting particle neural network (hereafter referred to as HIP-NN potential) and applied to predict the general properties of UN, its application to thermal modeling remains unclear. In this work, we specifically target the accurate prediction of bulk thermal conductivity in UN by combining the developed MTP with MD simulations and the relaxation time approximation (RTA) of the Boltzmann transport equation (BTE) for thermal conductivity calculation \cite{McGaughey2004,Srivastava2022}. Note that for materials with a wide phonon band gap like UN, at high temperatures, the RTA approach requires an evaluation of higher-order interatomic force constants (IFCs, up to the fourth order), which traditionally necessitates computationally expensive DFT calculations for thousands of supercell structures based on the frozen phonon method \cite{Esfarjani2008}. By substituting DFT with an accurate MTP, we enable the use of larger supercells and significantly reduce the computational cost, thereby enhancing the prediction capability of thermal conductivity comparing to experimental data. Moreover, the developed MTP also enables MD prediction of other thermal properties, such as melting point, thermal expansion, and specific heat, and the results show strong agreement with experimental measurements, further validating the reliability of the developed potential for evaluating thermal properties.

\section{Methods}

\subsection{Moment tensor potential}

Moment tensor potential (MTP) is a local potential that describes a system based on moment tensor descriptors of the atomic environment in the system\cite{Shapeev2016}. The potential energy of an atomic system described by MTP is defined as a sum of the energies of atomic environments of the individual atoms,

\begin{equation}
    E_\text{MTP} = \sum_{i} V(n_i) 
\end{equation}
where $n_i$ stands for the local environment of atom $i$ and $V(n_i)$ represents its corresponding atomic energy contribution. The local environment is described by a set of structural descriptors that encode the relative positions of neighboring atoms within a cutoff sphere centered on the atom of interest. Theoretically, any continuous function describing local environments can be approximated by linear combinations of polynomial basis functions $\textbf{B}_\alpha$

\begin{equation}
    V_i=\sum_\alpha \xi_\alpha \textbf{B}_\alpha
\end{equation}
The coefficients $\xi_\alpha$ corresponding to the basis functions $\textbf{B}_\alpha$ are optimized during the ML training process by fitting to the reference dataset, until the minimization of the objective function is achieved.
\begin{equation}
\sum_{k=1}^N\left[w_{\mathrm{e}}\left(E_k^{\mathrm{MTP}}-E_k^{\mathrm{DFT}}\right)^2+w_{\mathrm{f}} \sum_{i=1}^{N^{(k)}}\left|\mathbf{f}_{i, k}^{\mathrm{MTP}}-\mathbf{f}_{i, k}^{\mathrm{DFT}}\right|^2\right. \\ \left.+w_{\mathrm{s}} \sum_{a, b=1}^3\left(\sigma_{a b, k}^{\mathrm{MTP}}-\sigma_{a b, k}^{\mathrm{DFT}}\right)^2\right] \rightarrow \min 
\end{equation}

where $N$ is the number of configurations in the training dataset, $k$ is the index of each configuration and $i$ is the index of each atom within a configuration. $E$, $f$ and $\sigma$ denote energy, force and stress, respectively. The weights assigned to $E$, $f$ and $\sigma$, namely $w_{\mathrm{e}}=0.85$,$w_{\mathrm{f}}=0.15$ and $w_{\mathrm{s}}=0.001$ were determined after fine-tuning on the training dataset in this work.


\subsection{ML training procedure}

\begin{figure}[htbp]
    \centering
    \includegraphics[width=0.88\linewidth]{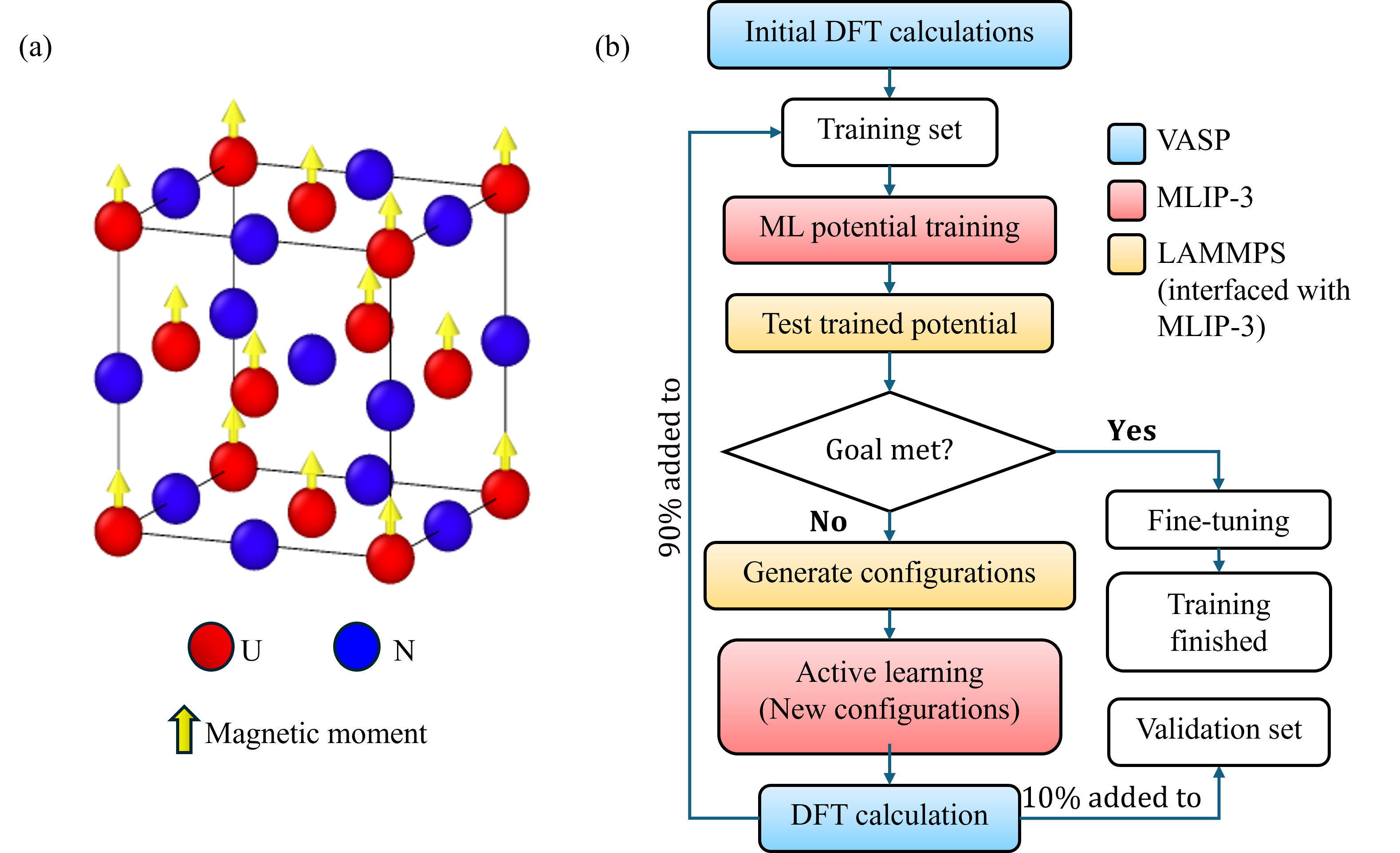}
    \caption{(a) Model of FM ordering of the spins of uranium atom. (b) Flowchart of the MLIP training procedure.}
    \label{fig:flowchart}
\end{figure}

For the generation of training set, all DFT calculations were performed using the Vienna $\textit{ab initio}$ simulation package (VASP 6.4.1) \cite{vasp1Kresse1996, vasp2Kresse1996}. The results presented were obtained using the generalized gradient approximation (GGA) functional for exchange correlation within the Perdew-Burke-Ernzerhof (PBE) functional \cite{Perdew1996}. In previous works, Hubbard U correction \cite{Claisse2016,Lan2013} has been applied to better account for the correlated nature of uranium’s 5f electrons. However, the systematically assessment of DFT for UN by Kocevski et al. \cite{Kocevski2022_spin} suggests that the use of Hubbard U correction causes imaginary phonons and negative C$_{44}$ elastic constants, which indicate dynamical unstability, while ferromagnetic (FM) ordering (as shown in Fig. \ref{fig:flowchart}a), combined with spin-orbit coupling (SOC), yields good agreement with experimental data and also shows dynamical stability without the use of Hubbard U. A further motivation for using FM ordering is consistency with prior studies \cite{Kocevski2022,alzate,Szpunar2020} to have a meaningful comparison of existing results. DFT calculations were performed using a conventional 2$\times$2$\times$2 supercell (64 atoms), and additional tests with 3$\times$3$\times$3 supercell (Table S1, Supplementary Material (SM) ) confirmed that finite-size effects on total energies and defect energetics are minor. With a much-reduced computational cost, the 2$\times$2$\times$2 cell represents a reasonable and practical choice. Hence, the PBE functional with SOC and FM ordering was employed, together with a plane-wave energy cutoff of 520 eV and Gaussian smearing with a width of 0.05 eV. Energy convergence criteria was set to $10^{-4}$ eV, and a 3$\times$3$\times$3 K-point mesh was employed for Brillouin zone sampling which has been verified to be sufficiently accurate (see SM Figure S1 and S2 for the convergence test). It is also important to note that the choice of functionals and lack of Hubbard U may cause inaccuracy for the C$_{11}$ and C$_{12}$ elastic constants calculation, which will be discussed later in Section \ref{physicalproperties}.


The overall ML training procedure is illustrated in Fig.\ref{fig:flowchart}b. It began with the generation of an initial training dataset from DFT calculations, which provide energies, forces, and stress tensors for a set of representative atomic configurations with different atomic displacements and lattice parameters. This dataset was used to train an initial MTP, and the developed MTP was then applied in MD simulations to predict energies and forces for various UN configurations. To further enhance the capability of the MTP, we employed it to generate new configurations via MD simulations at elevated temperatures up to 3600 K to obtain amorphous structures, and an active learning algorithm \cite{PODRYABINKIN2017171,Podryabinkin2023} selects representative structures to expand the training dataset and improve its diversity. DFT calculations were then performed on the selected configurations. The resulting data were randomly divided into a training set (90\%) and a validation set (10\%). A new MTP was subsequently refitted based on the previous MTP and the updated training set. This learning-on-the-fly iterative process continued until the MTP reached the target accuracy and the final dataset contains 3236 atomic configurations. This workflow is enabled by the MLIP-3 package \cite{Podryabinkin2023}, which constructs the MTP based on DFT-derived atomic configurations. The LAMMPS simulation package \cite{LAMMPS}, interfaced with MLIP-3 via the LAMMPS-MLIP-3 plugin \cite{novikov_interface_lammps_mlip_3}, is employed to integrate the trained potential into MD simulations.

\subsection{Property evaluation}
\subsubsection{Physical properties}

\noindent \textit{A.Equation of state (EOS)}

Strains were applied in the x, y, and z directions of a conventional 2$\times$2$\times$2 UN supercell. The potential energies of the strained structures were calculated using both MD with the developed MTP and DFT employing the same settings used in database generation. The lattice constants were determined by fitting of the energy-volume curves from both MTP and DFT  using the Birch-Murnaghan model \cite{murnaghan}.

\vspace{1em}
\noindent \textit{B.Elastic constants}

The cubic crystal structure of UN has only three independent elastic constants: $C_{11}$, $C_{12}$, and $C_{44}$. These constants were calculated using the energy–strain method \cite{mehl1994,Gueddim2010} based on a set of three equations. Starting from the equilibrium structure of the conventional 2$\times$2$\times$2 UN supercell obtained from the EOS, strains were applied simultaneously along $\pm x$, $\pm y$, and $\pm z$; along $\pm x$ and $\pm y$; and in the $\pm xy$ shear direction. The resulting energy–volume curves were then fitted to the corresponding equations to extract the elastic constants.

\vspace{1em}
\noindent \textit{C. Phonon dispersion}

We performed phonon dispersion calculations using the Phonopy package \cite{phonopy-phono3py-JPCM}, which employs the finite displacement method to compute second-order force constants \cite{phonopy-phono3py-JPSJ}. These calculations were conducted for both MTP and DFT using the conventional 2$\times$2$\times$2 UN supercell, utilizing a 10$\times$10$\times$10 k-point mesh grid to sample the Brillouin zone. The results were then compared to validate the accuracy of the MTP predictions.

\vspace{1em}
\noindent \textit{D. Defect formation energies}

To facilitate comparison with previous studies, the formation energy $E_\mathrm{f}$ of stoichiometric defects in UN, including Frenkel pairs (FPs) and Schottky defects (SDs), was predicted by MD calculations enabled by the developed MTP, with validation from DFT calculations. Both bound and unbound configurations were considered for uranium FP (U$_\mathrm{FP}$) and nitrogen FP (N$_\mathrm{FP}$) and Schottky defects. The formation energies are computed using the following expressions:
 
\begin{equation}
\begin{aligned}
    E_\mathrm{f}(\text{FP}) &= E_\mathrm{d} - E_\mathrm{p} \\
    E_\mathrm{f}(\text{SD})&= E_\mathrm{d} - \frac{N-2}{N}E_\mathrm{p} \\
\end{aligned}
\label{eq:Ef}
\end{equation}
where $N$ is the total number of atoms in the perfect supercell, $E_\mathrm{p}$ is its total energy, and $E_d$ is the potential energy of the defect-bearing supercell of the same size. Calculations are performed using the conventional 2$\times$2$\times$2  supercell of UN. Individual point defect is generated by removing or inserting U/N into the initially perfect cell. Bound configurations (FP or SD) are constructed by introducing both defects (a vacancy and an interstitial, or two vacancies) within a single supercell, whereas unbound configurations place the defects in separate supercells to simulate spatial separation.

\subsubsection{Thermal properties}

\noindent \textit{A. Thermal expansion}

For thermal expansion, we computed the relative linear thermal expansion (rLTE), defined as,
\begin{equation}
    \text{rLTE} = \frac{a(T)-a_{300}}{a_{300}} 
    \label{eq:rLTE}
\end{equation}
where $T$ is the temperature in Kelvin, $a(T)$ is the equilibrium lattice parameter at temperature $T$, and $a_{300}$ denotes the lattice parameter at 300 K. 

To determine $a(T)$, MD simulations were conducted in the isothermal-isobaric (NPT) ensemble maintained at isotropic zero external pressure with the temperature range from 300 K to 1800 K in increments of 50 K, using the Nose–Hoover thermostat and barostat\cite{Nos1984,Hoover1985}. For each temperature, the system was equilibrated for 20 ps and the box dimensions were recorded every 1 ps and averaged to obtain the corresponding lattice parameter. Each simulation used a conventional $8\times 8\times 8$ UN supercell (See SM Figure S3 and S4 for the convergence test on supercell size) of the UN unit cell, comprising 4096 atoms. The average lattice parameter at each temperature was obtained by averaging the box dimensions. A timestep of 1.0 fs was used for all simulations in this work. The calculated $a(T)$ values were then used in Eq. \ref{eq:rLTE} to determine rLTE.

\vspace{1em}
\noindent \textit{B. Melting point}

The conventional 8$\times$8$\times$8 UN  supercell was also used for the MD simulations for melting point. The system was heated under the NPT ensemble with the temperature ramped from 2600 K to 3600 K and isotropic zero external pressure using the Nose–Hoover thermostat and barostat\cite{Nos1984,Hoover1985}. A timestep of 1 fs was used throughout the simulation. The enthalpy per atom was computed during the run and averaged every 4 ps, and the entire simulation was carried out over 200 ps. In this work, the melting point of UN was estimated from the temperature at which the crystal becomes mechanically unstable in MD simulations. This approach provides only an approximate value, as the true melting point requires more rigorous methods such as the moving interface approach \cite{Govers2008}. Moreover, experiments and thermodynamic models \cite{Chevalier2000,Kocevski2022} show that UN decomposes into a gas and a liquid rather than undergoing congruent melting, which cannot be reproduced by the current ML potential. Our results should therefore be interpreted as the onset of crystalline instability rather than a rigorous melting point.


\vspace{1em}
\noindent \textit{C. Specific heat}

The specific heat capacity at constant zero pressure $C_\mathrm{p}$ of UN is calculated by
\begin{equation}
    C_\mathrm{p} = \frac{\partial{H}}{\partial{T}} 
    \label{eq:specificheat}
\end{equation}

where $H$ is the enthalpy of system in J/mol. The MD simulations for $C_\mathrm{p}$ have also used the  conventional 8$\times$8$\times$8 UN supercell under a NPT ensemble at isotropic zero external pressure with an extended temperature range from 250 K to 1850 K in increments of 50 K, using the Nose–Hoover thermostat and barostat\cite{Nos1984,Hoover1985}. For each temperature, the enthalpy of the system was obtained by averaging over 20 ps, with data collected at 1 ps intervals. The specific heat was then calculated over the range of 300 K to 1800 K using Eq. \ref{eq:specificheat}. In parallel, the heat capacity was also calculated using lattice dynamics (LD) within the quasi-harmonic approximation implemented in phonopy-qha \cite{Togo2010}. The calculations were performed with a conventional  2$\times$2$\times$2 UN supercell using both the developed MTP and the Kocevski potential.

\vspace{1em}
\noindent \textit{D. Thermal conductivity (Computation)}

The total thermal conductivity $\kappa$ of UN consists of both lattice thermal conductivity ($\kappa_{L}$) and electronic thermal conductivity($\kappa_\mathrm{e}$), i.e., $\kappa_\mathrm{total} = \kappa_\mathrm{L} + \kappa_\mathrm{e}$. In this work, $\kappa_{L}$ is calculated using the RTA approach,

\begin{equation}
    \kappa_\mathrm{L} = \Sigma c_V v^2 \tau
\end{equation}
where $c_V$ is the phonon specific heat at constant volume, $v$ is the phonon group velocity, and $\tau$ is the phonon relaxation time, which is evaluated via Matthiessen’s rule:
\begin{equation}
    \tau^{-1} = \tau_\mathrm{3ph}^{-1} + \tau_\mathrm{4ph}^{-1}+\tau^{-1}_\mathrm{b}+\tau^{-1}_\mathrm{iso}+\tau_\mathrm{elph}^{-1}
\end{equation}
where $\tau^{-1}$, the reciprocal of phonon relaxation time, represents phonon scattering rate. $\tau_\mathrm{3ph}^{-1}$ and $\tau_\mathrm{4ph}^{-1}$ denote the three-phonon and four-phonon scattering rates, respectively, which were calculated using the Phonopy and ShengBTE packages \cite{Han2022,guo2023} in this work. Forces were calculated with the developed MTP for displaced supercell configurations and the second-, third-, and fourth-order IFCs were extracted based on these configurations and forces. The scalebroad parameter was set 0.5 and mesh was set 30$\times$30$\times$30 in ShengBTE. A primitive 5$\times$5$\times$5 UN supercell containing 250 atoms, (see Figure S5 in SM for the convergence test) was used for phonon scattering rates calculation for $\kappa_{L}$. The boundary ($\tau^{-1}_\mathrm{b}$) and isotope ($\tau^{-1}_\mathrm{iso}$) phonon scattering rates are not the primary focus of this work. $\tau^{-1}_\mathrm{b}$ was computed using ShengBTE's default settings, while $\tau^{-1}_\mathrm{iso}$ was considered using the Tamura model as implemented in ShengBTE. Finally, the electron-phonon scattering rates, $\tau_\mathrm{elph}^{-1}$, were computed by combining electron data from DFT output of UN and phonon data from the developed MTP using the EPW code \cite{Lee2023,Ponce2016}. The electron-phonon matrix elements required for the EPW code were initially calculated on a coarse 4$\times$ 4$\times$ 4 q grid, and then interpolated onto ﬁne grids using maximally localized Wannier functions (MLWFs)\cite{Marzari2012} construct by random projection method \cite{Mostofi2014} to calculate electron-phonon scattering rates. The obtained $\tau_\mathrm{elph}^{-1}$ were incorporated into the total phonon scattering rates via Matthiessen’s rule for thermal conductivity calculations.

$\kappa_\mathrm{e}$ is evaluated using the Wiedemann–Franz law \cite{Franz1853}: 

\begin{equation}
    \kappa_\mathrm{e} = \frac{LT}{\rho} 
    \label{eq:kappa_e}
\end{equation}
where $L=2.44\times10^{-8}\mathrm{V^{2}K^{-2}}$ is the Lorenz number\cite{Lorenz1872}, $\rho$ is the electrical resistivity obtained from an empirical single spacing ($\delta=170$$\mathrm{K}$) crystal field (CF) model fitted based on experimental results \cite{Samsel}:
\begin{equation}
    \rho(T) = \rho_0 + \rho_{\mathrm{CF}}(\infty)/\mathrm{cosh}^{2}(\delta/2T) 
    \label{eq:rho}
\end{equation}
where $\rho_0=85.6$$\mathrm{\mu\Omega cm}$, $\rho_{\mathrm{CF}}(\infty)=63.3$$\mathrm{\mu\Omega cm}$.

We note that while phonon transport analysis provides detailed insight into phonon scattering processes, direct MD methods such as Green–Kubo \cite{Green1954,Kubo1957},or non-equilibrium molecular dynamics (NEMD) are more suitable at high temperatures or in the presence of defects, and will be explored in future work.

\vspace{1em}
\noindent \textit{E. Thermal conductivity (Experiment)}

The thermal conductivity of UN was obtained by measuring its thermal diffusivity and combining it with literature values of heat capacity and density. For the experimental measurements, a UN sample was prepared using the UN powder feedstock, synthesized via the carbothermic reduction method, received from Los Alamos National Laboratory. The as‑received powder was subjected to a high energy ball milling process (300 rpm, 1 hour, Retsch PM 200 planetary ball mill) in a ZrO$_2$ milling vessel with 5 mm diameter ZrO$_2$ media and the milled powder was then passed through a 400‑mesh (37 $\mu$m) sieve prior to pressing into pellets of right cylindrical geometry using an automated Carver hydraulic press at $\approx$670 MPa with an 8.25 mm WC die. A small amount of dry zinc stearate (Sprayon MR312) was used as a lubricant on the die surface, die punches, and die faces. All milling and pellet fabrication, and sintering activities occurred in an inert atmosphere glovebox having under 5 ppm of O$_2$.

The UN pellets were sintered in a refractory metal sintering furnace (Thermal Technology Model 1100) with pellets placed directly on a tungsten sintering plate within the hot zone of the furnace. A sintering cover gas of oxygen scrubbed Ar/+ 100 ppm N$_2$ (1 L/min) was used during sintering. The Ar/N$_2$ mixture is used to avoid UN dissociation and formation of U$_2$N$_3$ during the sintering ramp (25 °C/min), at the dwell temperature, 1950°C, for 25 hours, and during the cool‑down phase to 1200 °C \cite{TENNERY1971}. Once the sample reached 1200 °C, the atmosphere was switched to UHP Ar for the final cool-down until room temperature to prevent sesquinitride formation.

Thermal diffusivity was measured over the temperature range of 77 K to 300 K, using the spatial-domain thermoreflectance (SDTR) technique. In the SDTR system, an intensity-modulated 660-nm continuous wave (CW) laser is used to locally heat up the sample surface and generate the thermal wave (transient temperature variation in both time and spatial domains). The thermal wave propagation in the spatial range of 20 $\mu$m is probed by a 532-nm laser through thermoreflectance effect. The measurement probing range can be estimated by thermal diffusion length, given by $L_{\mathrm{th}}=\sqrt{D/\pi/f}$, where $D$ is the thermal diffusivity and $f$ is the heating laser modulation frequency. In this study, the heating laser is modulated in a frequency range of 20-100KHz, corresponding to thermal diffusion lengths of 4.5-10 $\mu$m (assuming $D\approx 6 \mathrm{mm^2/s}$ for UN). As this length scale is significantly smaller than the grain diameter (50-100 $\mu$m), the measurements predominantly probe intra-grain heat transport and can be considered representative of single-crystal behavior.

Both the heating and probe laser beams were focused to $\sim$1.5 $\mu$m-radius spots on the surface of the sample using a 50$\times$ objective lens, and the optical power at the sample surface was $\sim$3 mW and $\sim$0.3 mW for the heating and probe lasers, respectively. In order to ensure a sufficient thermoreflectance response, samples were coated with a thin layer of gold as a transducer \cite{Hurley2015}. A relatively small thickness ($\sim$30 nm) was selected to reduce the impact of film properties on measurements \cite{Hua2012}. Thermal properties of the deposited gold films were determined from co-deposited films of the same thickness on pristine NBK7 substrates. The change of the gold film thermal properties with respect to temperature were considered through an empirical approach \cite{Yurgens2010}. The details of this procedure can be found in previous publications \cite{Deskins2022,Dennett2020,Hua2020}.

\section{Results and Discussion}

\subsection{MLIP training}

\begin{figure}[htbp]
    \centering
    \includegraphics[width=0.9\linewidth]{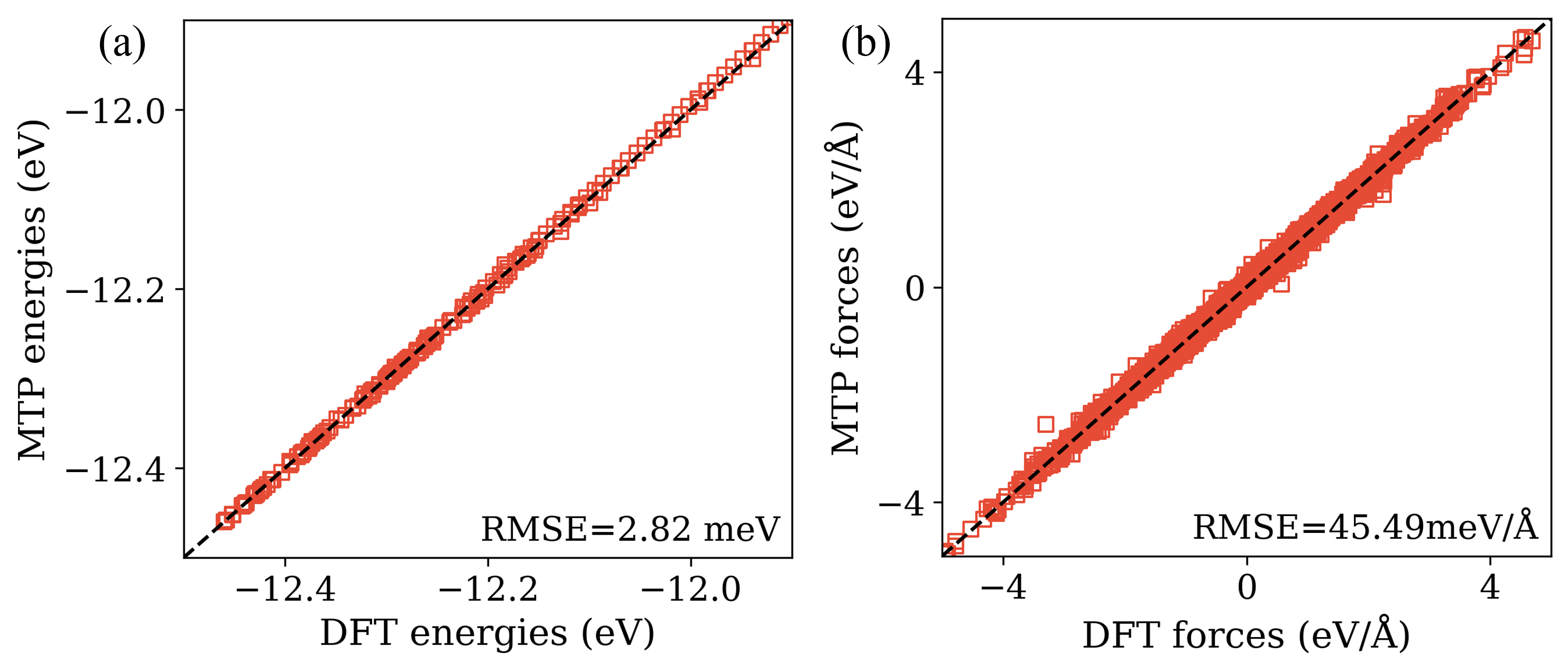}
    \caption{(a) Energies and (b) forces predicted by the developed MTP compared to the DFT reference data in validation set for the corresponding conventional 2$\times$2$\times$2 UN supercells.}
    \label{fig:E_f_rmse}
\end{figure}

The accuracy of the developed MTP was evaluated by comparing its predictions of energies and forces against the DFT reference values in the validation set which includes 450 atomic configurations. The error was quantified via the root-mean-square error (RMSE). As shown in Figures \ref{fig:E_f_rmse}a and \ref{fig:E_f_rmse}b, the RMSE for MTP-predicted energies and forces relative to DFT values are 2.82 meV/atom and 45.49 meV/Å, respectively. These low RMSE values indicate good agreement between the MTP and DFT calculations. As a reference, Alzate-Vargas et al. obtained 2.48 meV/atom and 111 meV/Å in HIP-NN for UN material \cite{alzate}.  Hence, this level of accuracy achieved in this study ensures that the MTP is suitable not only for static property predictions (e.g., defect formation energies, elastic constants) but also for temperature-dependent simulations such as thermal transport and thermal expansion.  

\subsection{Physical properties}
\label{physicalproperties}
 
\begin{figure}[htbp]
    \centering
    \includegraphics[width=0.68\linewidth]{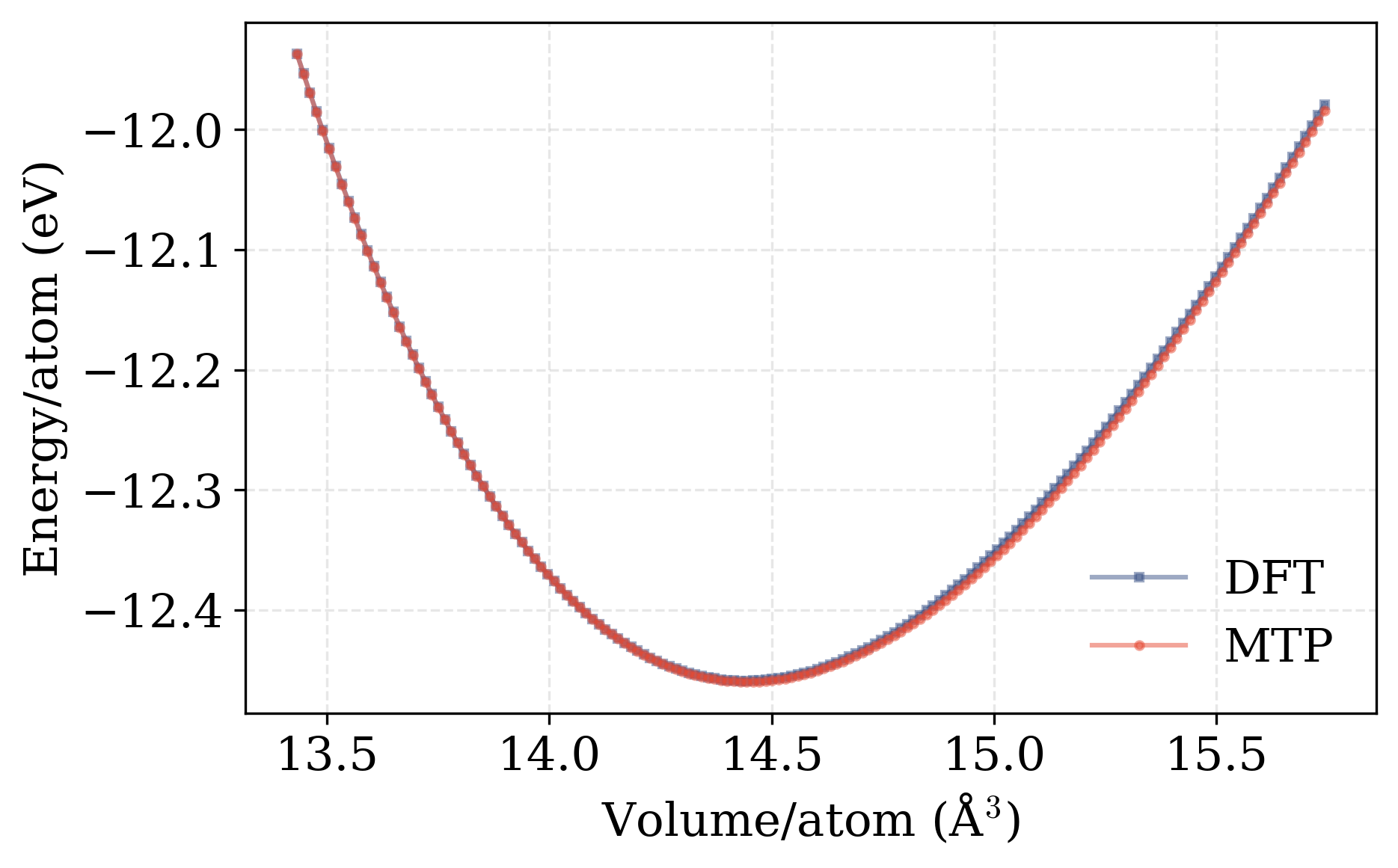}
    \caption{ Equation of state predicted by the developed MTP, compared with DFT results calculated using a conventional 2$\times$2$\times$2 UN supercell with the same settings as those employed for database generation. }
    \label{fig:eos}
\end{figure}

The equation of state results presented in Fig. \ref{fig:eos} compares the energy–volume curves obtained from both MTP and DFT calculations. The two methods yielded very consistent values across the whole volume range, which implies that the developed MTP accurately captures the underlying cohesive energy landscape of UN. The lattice constants ($a_0$) extracted from the energy-volume curves in Fig. \ref{fig:eos} is listed in Table \ref{tab:alattice} and compared with the values from Tseplyaev, Kocevski and HIP-NN potentials \cite{AbdulHameed2024,alzate}. Among all the potentials considered, the lattice constant predicted by the developed MTP shows the closest agreement with the experimental value at 300 K. In contrast, the Tseplyaev potential exhibits a noticeable overestimation of the lattice constant compared to the other potentials.

Accurate prediction of elastic properties is essential for modeling the mechanical behavior of UN under operational and extreme conditions. Starting from the equilibrium structure, the three independent elastic constants C$_{11}$, C$_{12}$ and C$_{44}$ were calculated from energy–strain method \cite{mehl1994,Gueddim2010}, which used molecular statics calculations of strained supercells to extract those quantities. From these, the Bulk modulus $B$, Young’s modulus $Y$ and Shear modulus $G$ were derived using the Voigt-Reuss-Hill (VRH) averaging scheme \cite{Hill1952,Anderson1963}. Both MTP- and DFT- predicted and elastic constants and moduli are compared with both experimental data and calculation results from previous studies \cite{AbdulHameed2024, alzate, Salleh1986, Hayes1990gp} in Table \ref{tab:alattice}. Since an MLIP can only reproduce the accuracy of its DFT reference data, discrepancies between the predicted elastic constants and experimental data reflect the inherent limitations of DFT calculation itself. As shown in Table \ref{tab:alattice}, different DFT functionals and setups can significantly affect the calculated values of C$_{11}$, C$_{12}$ and C$_{44}$ \cite{alzate}. The PBE+U gives C$_{11}$ and C$_{12}$ values that are very close to experiment, but predicts a negative C$_{44}$. PBE gives better C$_{11}$ than PBE+SOC, but worse C$_{12}$ and similar C$_{44}$. The SCAN functional gives better C$_{12}$ than PBE, but much worse C$_{11}$ and C$_{44}$. 

\begin{table}[htbp]
\begin{tabular*}{\textwidth}{@{\extracolsep{\fill}}llllllll}
\hline
\rule{0pt}{13pt} 
              & $a_0$ (\AA) & C$_{11}$   & C$_{12}$    & C$_{44}$   & $B$     & $E$     & $G$    \\
              \hline
MTP (this work) & 4.87 & 379.2 & 111.9  & 51.2  & 210.0 &  205.7 & 77.5 \\
DFT PBE+SOC (this work)  & 4.87 & 390.5 & 99.7 & 52.9  & 196.6 & 212.33 & 80.4 \\
DFT PBE  \cite{alzate}       &  & 405   & 127    & 52    &   &       &      \\
DFT PBE+U \cite{alzate}      &  & 416   & 102    & -18    &    &       &      \\
DFT SCAN \cite{alzate}       &  & 368   & 111    & 33    &    &       &  \\
Tseplyaev  \cite{AbdulHameed2024}    & 4.81 & 586.6 & 110.5  & 54.7  & 269.2 & 260   & 105  \\
Kocevski  \cite{AbdulHameed2024}     & 4.90 & 425.4 & 117.0  & 71.0  & 219.8 & 250   & 95   \\
HIP-NN   \cite{alzate}      & 4.86* & 393   & 131    & 43    & 221   &       &      \\
Experiment    & 4.88 & 423.9* & 98.1*   & 75.7*  & 206.7* & 240* & 105*  \\
\hline
\end{tabular*}
(* at 300 K)
\caption{Lattice constant $a_0$ (in \AA), elastic constants C$_{11}$, C$_{12}$, C$_{44}$, bulk modulus $B$, Young's modulus $E$, and shear modulus $G$ (all in GPa) from the developed MTP potential and DFT calculations using a conventional 2$\times$2$\times$2 UN supercell, compared with DFT values using PBE, PBE+U and SCAN functionals (taken from Ref. ~\cite{alzate}), MD values from the Tseplyaev, Kocevski, and HIP-NN potentials (taken from Ref. ~\cite{AbdulHameed2024,alzate}), and experimental values taken from Ref. ~\cite{Salleh1986,Hayes1990gp,Marples1970}. All DFT calculations are using FM ordering in UN structures.}
\label{tab:alattice}
\end{table}

\begin{figure}[htbp]
    \centering
    \includegraphics[width=0.75\linewidth]{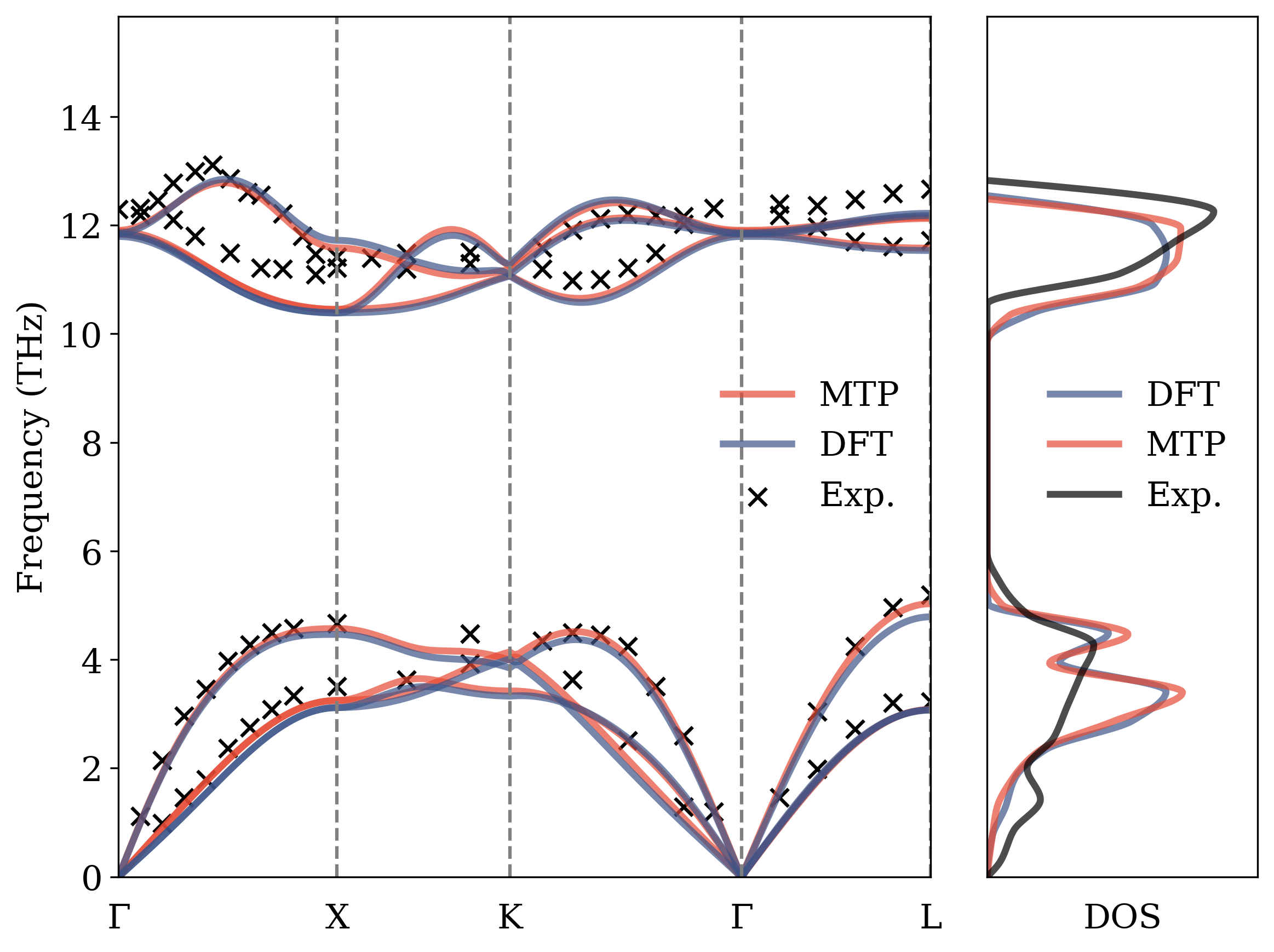}
    \caption{Phonon dispersion and density of state (DOS) of UN predicted by the trained MTP and DFT calculations using a conventional 2$\times$2$\times$2 UN supercell with 10$\times$10$\times$10 mesh grid, compared with the experimental data taken from Jackman et al. \cite{Jackman1986}.}
    \label{fig:phonon_dispersion}
\end{figure}

We next assessed the ability of the MLIP to capture near-equilibrium vibrational properties by examining the phonon dispersion of UN. Phonons play a key role in thermal transport, particularly in materials with non-negligible contribution to thermal transport via phonons. The phonon dispersion relations calculated using the developed MTP are shown in Figure \ref{fig:phonon_dispersion}, along with those derived from DFT calculations and experimental data reported by Jackman et al. \cite{Jackman1986}. Overall, the MTP- and DFT-predicted phonon dispersions show excellent agreement, with nearly overlapping curves across the entire Brillouin zone. The MLIP captures the main features of the experimental dispersion, particularly in the acoustic branches, while some discrepancies are observed in the optical branches. Despite this, the strong agreement in the acoustic regime is encouraging; to provide reasonable prediction of $\kappa_L$, as low-frequency acoustic phonons typically dominate lattice thermal conductivity due to their high group velocities and longer lifetimes.

Finally, we evaluated the defect formation energies ($E_f$) of stoichiometric defects in UN. This serves a critical assessment of the potential to be applied to radiation damage studies of UN as nuclear fuel.  The computed values of $E_f$ for bound and unbound U$_\text{FP}$, N$_\text{FP}$ and SD are summarized in Table \ref{tab:defect}, along with corresponding results from previously developed interatomic potentials, and the DFT calculations using the same settings as those employed for database generation. The MTP-predicted defect formation energies are in good agreement with DFT benchmarks. By comparison, a notable discrepancy was observed in the predictions from the Kocevski potential, which significantly overestimates the formation energies of both bound and unbound U\textsubscript{FP}. This deviation has been previously attributed to deficiencies in the Kocevski potential’s description of metallic uranium, which affect the accuracy of defect energetics in uranium-rich or stoichiometric environments \cite{AbdulHameed2024}. In contrast, the MTP trained in this work demonstrates improved transferability and accuracy across multiple defect configurations, which supports its suitability for radiation damage modeling in UN.

\begin{table}[htbp]
    \centering
    \begin{tabular*}{\textwidth}{@{\extracolsep{\fill}}lllllll}
    \hline
    \rule{0pt}{13pt} 
         Defects $E_\mathrm{f}$ (eV) &  MTP & DFT & Tseplyaev & Kocevski & HIP-NN \\
         \hline
        unbound U$_\text{FP}$ & 9.04 & 9.83 & 9.32 & 14.41 & 9.86 \\
        bound U$_\text{FP}$   & 8.73 & 9.04 & 7.26 & 10.30 &  \\
        unbound N$_\text{FP}$ & 4.21 & 4.59 & 4.67 & 4.03  & 4.99 \\
        bound N$_\text{FP}$   & 4.04 & 4.55 & 3.76 & 3.23  & \\
        unbound SD  & 3.88 & 4.76 & 4.57 & 3.98  & 5.16 \\
        bound SD    & 3.30 & 4.57 & 4.51 & 3.99  & 4.7 \\
        \hline
    \end{tabular*}
    \caption{
    Formation energies ($E_\mathrm{f}$) of bound and unbound U$_\text{FP}$, N$_\text{FP}$ and Schottky defects predicted by MTP and DFT calculations using a conventional 2$\times$2$\times$2 UN supercell. Reference values of $E_\mathrm{f}$ from the Tseplyaev and Kocevski potentials are taken from AbdulHameed et al. \cite{AbdulHameed2024}, while those from the HIP-NN potential are taken from Alzate-Vargas et al. \cite{alzate}}
    \label{tab:defect}
\end{table}


\subsection{Thermal properties}

\subsubsection{Thermal expansion}
\begin{figure}[htbp]
    \centering
    \includegraphics[width=0.98\linewidth]{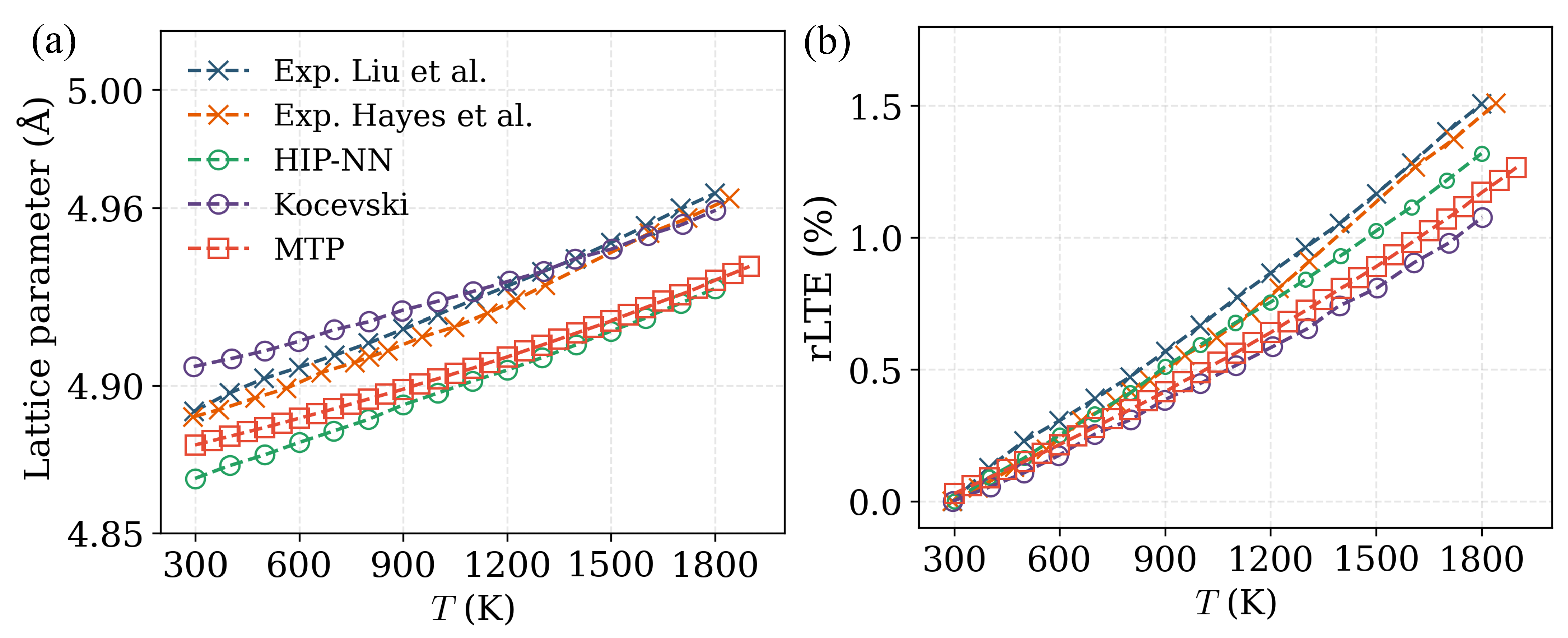}
    \caption{(a) Lattice parameter and (b) relative linear thermal expansion (rLTE in \%)  as functions of $T$ predicted by MTP using MD simulation with a conventional 8$\times$8$\times$8 UN supercell, compared with experimental results from Liu et al. (data points taken from  Ref ~\cite{LIU2023154215}) and Hayes et al. (data points taken from Ref ~\cite{Hayes1990}) as well as MD simulation results obtained using the HIP-NN potential (data points taken from Ref. ~\cite{alzate}) and the Kocevski potential (data points taken from Ref.~\cite{Kocevski2022}). }
    \label{fig:expansion}
\end{figure}

Thermal expansion is a fundamental property of materials that reflects the anharmonicity of atomic interactions and  and plays a critical role in applications involving dimension stability and thermal stress. The lattice parameter as a function of temperature is illustrated in Fig. \ref{fig:expansion}a, and the rLTE, calculated using Eq. \ref{eq:rLTE}, is presented in Fig. \ref{fig:expansion}b.  These results are compared against experimental measurements by Liu et al. \cite{LIU2023154215} and Hayes et al. \cite{Hayes1990}, as well as MD predictions using the HIP-NN potential \cite{alzate} and the empirical Kocevski potential \cite{Kocevski2022}. The MTP-predicted lattice parameter increases smoothly with temperature and exhibits a thermal expansion trend that closely follows the experimental data. However, a slight overall deviation in the absolute lattice constant was observed across the full temperature range. Both MLIPs (MTP and HIP-NN) yield similar predictions, MTP shows slightly better agreement with experimental values at lower temperatures, while the Kocevski empirical potential shows slightly better agreement with experimental values at high temperatures. Comparing the rLTE values, all potentials yield very consistent values compared to the experiments, although a slight underestimation is observed for all potentials at higher temperatures. Overall, those potentials and experimental measurements demonstrate good consistency in capturing the thermal expansion behavior of UN.

\subsubsection{Specific heat}
\begin{figure}[htbp]
    \centering
    \includegraphics[width=0.75\linewidth]{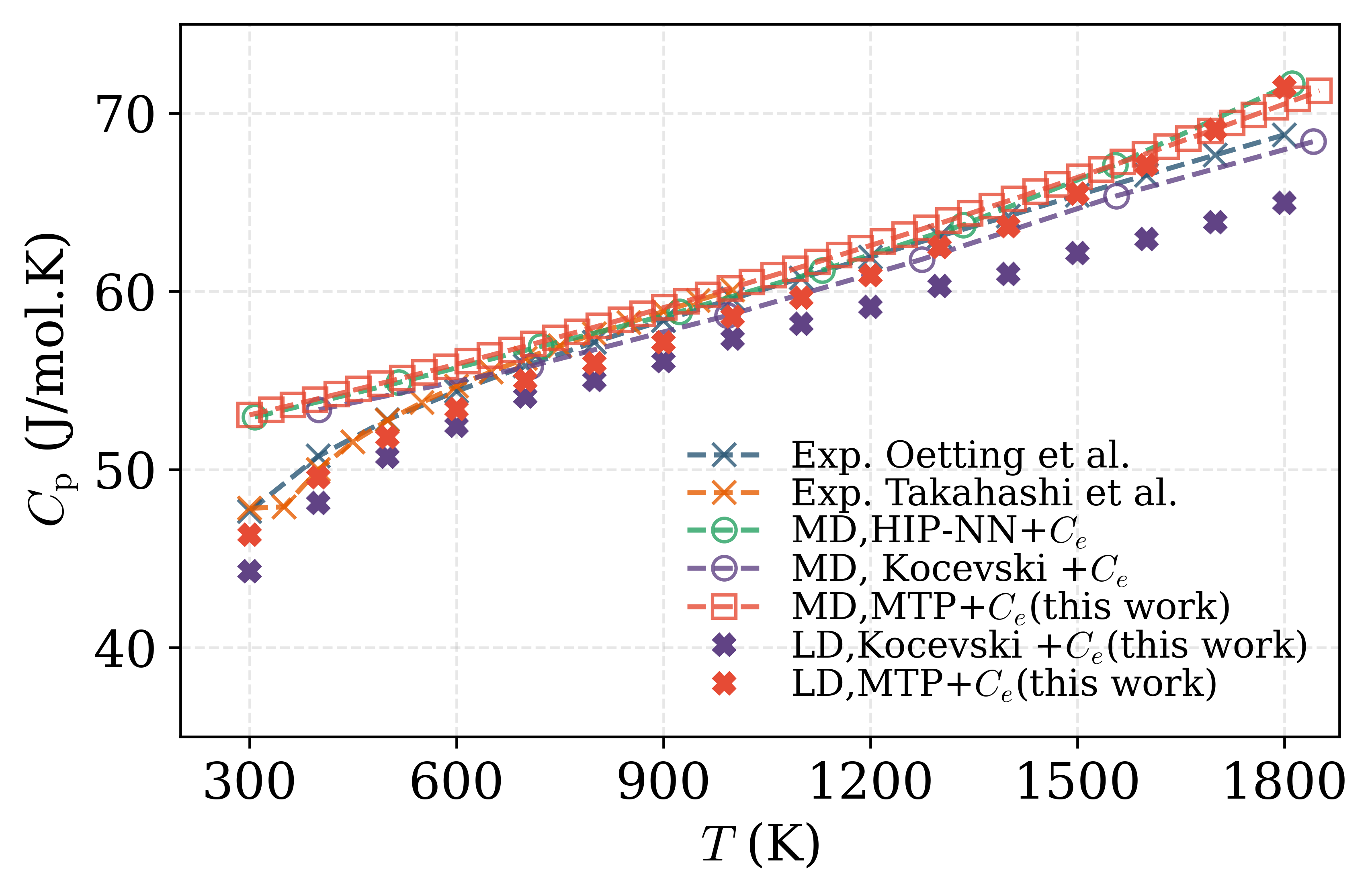}
    \caption{Specific heat of UN with respect to temperature predicted using MD simulations with the developed MTP and lattice dynamics (LD) calculations the developed MTP and Kocevski potential. A conventional 8$\times$8$\times$8 UN supercell was used for MD simulations and a conventional 2$\times$2$\times$2 UN supercell was used for LD calculations. The experimental data are taken from Oetting et al. \cite{Oetting1972} and Takahashi et al. \cite{Takahashi1971}. The molecular dynamics (MD) results obtained using HIP-NN are taken from Alzate-Vargas et al. \cite{alzate}, while the MD results using the Kocevski potential are taken from Kocevski et al. \cite{Kocevski2022}.   }
    \label{fig:specificheat}
\end{figure}

In addition to thermal expansion, MD simulations with MTP and LD calculations with both MTP and Kocevski potential are also employed to calculate the specific heat ($C_p$) of UN using the developed MTP, under the same conditions as those used for the thermal expansion analysis. The results, presented in Fig. \ref{fig:specificheat}, are compared with MD simulation data obtained using the HIP-NN and Kocevski potentials \cite{alzate,AbdulHameed2024}, as well as with experimental measurements reported by Takahashi et al. \cite{Takahashi1971} and Oetting et al. \cite{Oetting1972}. Additionally, the electronic contribution to the specific heat of FM UN, calculated from DFT by Szpunar et al. \cite{Szpunar2020}, is included in all MD and LD results to ensure consistency with experimental data.
 
At temperatures above 600 K, the MD and LD predictions by MTP show excellent mutual consistency and good agreement with experimental data. However, at lower temperatures (below 600 K), a consistent overestimation of specific heat is observed across MD results with all three potentials, with the deviation becoming more pronounced as the temperature decreases. This is because classical MD simulations are known to systematically overestimate heat capacity at low and intermediate temperatures because they treat atomic vibrations within the equipartition approximation, where each phonon mode contributes $k_B$T regardless of frequency\cite{vanGunsteren2024}. This neglects the Bose–Einstein statistics of phonon occupation, which strongly suppresses the effect of high-frequency modes near room temperature \cite{Baroni2010}. To provide a more rigorous comparison, we calculated the heat capacity using lattice dynamics within the quasi-harmonic approximation. These LD calculations, based on both our trained MTP potential and the Kocevski potential \cite{Kocevski2022}, yield results much closer to experiment than classical MD at low temperature. Furthermore, the MTP potential provides better agreement with experimental data than the Kocevski potential, demonstrating its improved predictive capability.

\subsubsection{Melting Point}

\begin{figure}[htbp]
    \centering
    \includegraphics[width=0.6\linewidth]{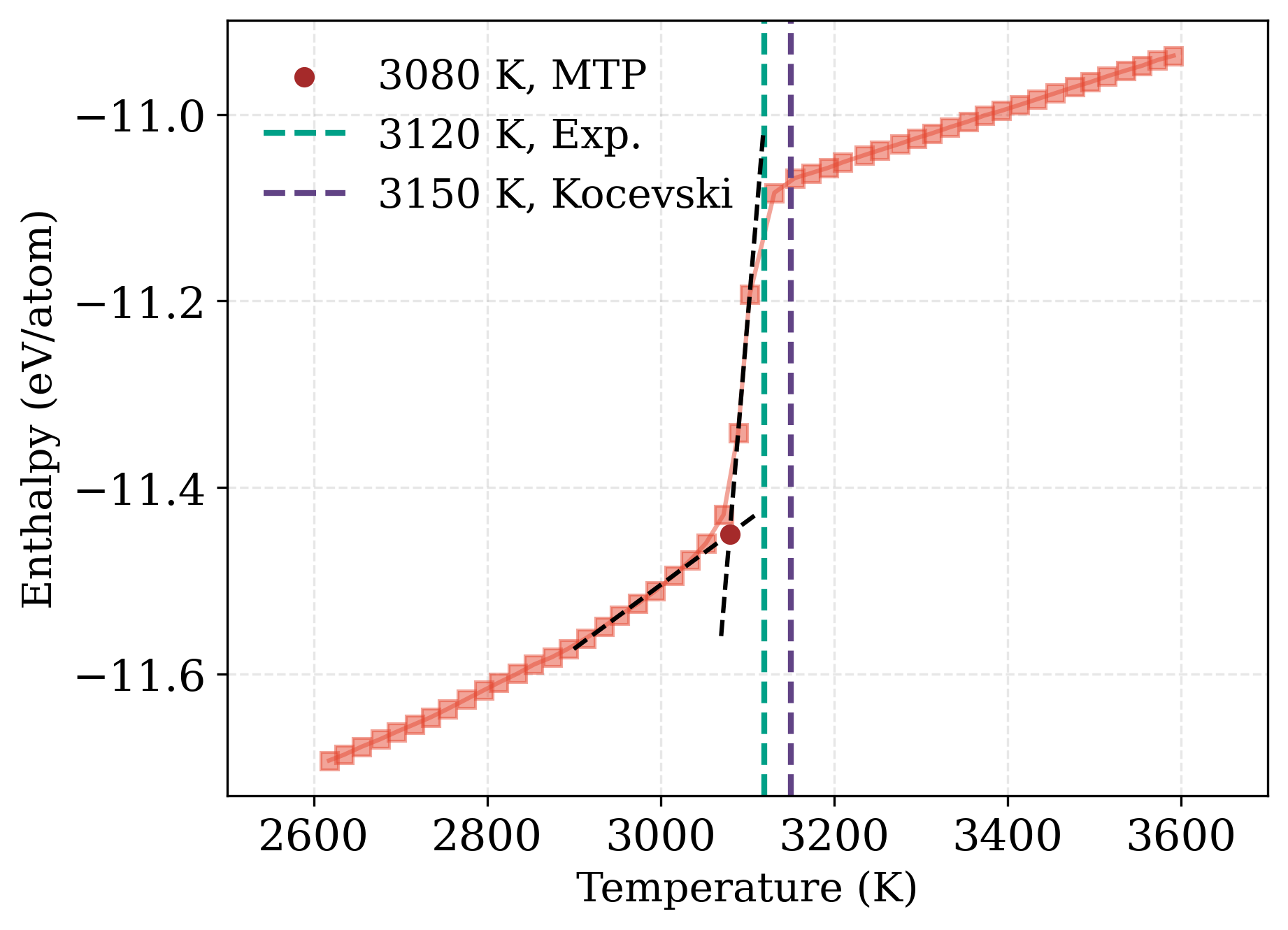}
    \caption{ Prediction of the melting point of UN via the enthalpy-temperature relationship using the developed MTP with MD simulation using a conventional 8$\times$8$\times$8 UN supercell. The predicted melting point value is compared with experimental data (taken from Ref. ~\cite{CARVAJALNUNEZ20141}) and MD data using Kocevski potential ( taken from Ref. ~\cite{AbdulHameed2024}). }
    \label{fig:melting}
\end{figure}

Here, the melting point $T_\mathrm{m}$ of UN was calculated using MD simulation.  As shown in Fig. \ref{fig:melting}, the enthalpy as a function of temperature was plotted to identify the onset of melting. The melting point corresponds to a sharp increase in enthalpy, with the latent heat associated with the solid–liquid phase transition. The exact value of  $T_\mathrm{m}$ is determined by the intersection of the two linear fits (one applied to the solid-phase enthalpy data and the other to the transition region), as shown in Figure \ref{fig:melting}. From this analysis, $T_\mathrm{m}$ is estimated to be approximately 3080 K. This value shows excellent agreement with the experimentally recommended melting point of 3120$\pm$30 K \cite{CARVAJALNUNEZ20141} and is also close to the 3150 K predicted using the Kocevski potential \cite{AbdulHameed2024}. As noted in the Methods, the reported melting point reflects crystalline instability in MD and should be regarded as an approximate value. Nevertheless, the MLIP remains robust for capturing high-temperature behavior and phase instability.

\subsubsection{Thermal conductivity}
\label{kappa}

\begin{figure}[htbp]
    \centering
    \includegraphics[width=0.98\linewidth]{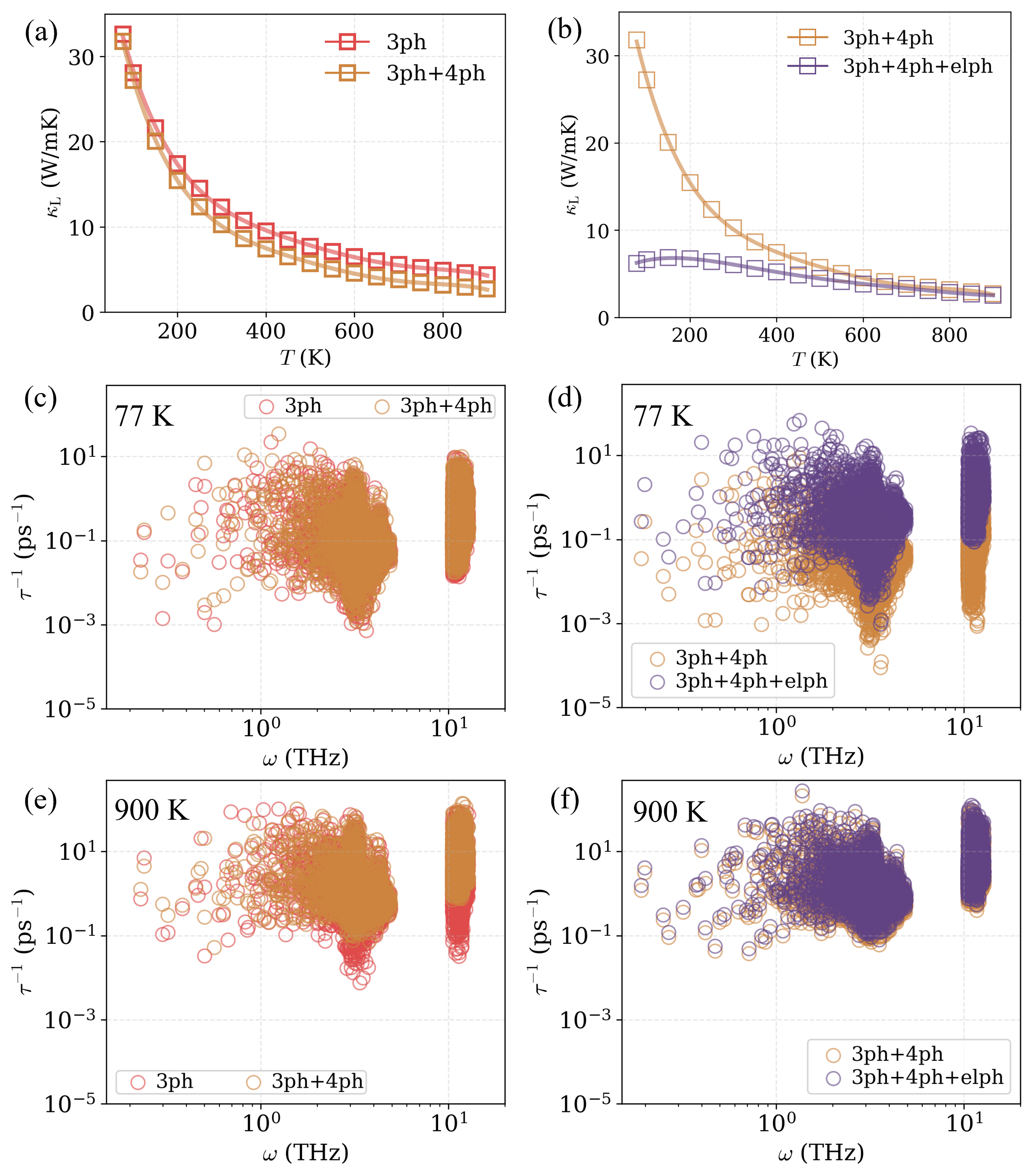}
    \caption{(a) Comparison of lattice thermal conductivity ($\kappa_\mathrm{L}$) considering only three-phonon scattering and both three-phonon and four-phonon scattering from 77 to 900 K. (b) Effect of electron-phonon scattering on $\kappa_\mathrm{L}$ over the temperature range of 77–900 K. (c) Comparison of phonon scattering rates ($\tau^{-1}$) considering only three-phonon scattering and both three-phonon and four-phonon scattering at 77 K. (d) Effect of electron-phonon scattering on $\tau^{-1}$ at 77 K. (e) Comparison of $\tau^{-1}$ considering only three-phonon scattering and both three-phonon and four-phonon scattering at 900 K. (f) Effect of electron-phonon scattering on $\tau^{-1}$ at 900 K. Calculations are based on the primitive 5$\times$5$\times$5 UN supercell. }
    \label{fig:kappa_scatter}
\end{figure}

To evaluate how well the developed MTP predicts lattice thermal conductivity ($\kappa_L$), we examined how different orders of phonon scattering affect the results. Previous studies include only three-phonon scattering when solving the BTE \cite{Szpunar2020}, but recent work in many other systems (e.g., \cite{Feng2017,Feng2018}) has shown that four-phonon scattering can become important, especially at high temperatures. In this work, we considered both three- and four-phonon scattering and directly compared their effects on the thermal conductivity of UN. Figure \ref{fig:kappa_scatter}a shows the total lattice thermal conductivity calculated with and without four-phonon scattering. At low temperatures, the effect of four-phonon scattering is negligible, while the impact of four-phonon scattering becomes substantial, leading to a noticeable reduction in $\kappa_L$ (e.g., 40 \% reduction at 900 K in this work). This trend is expected, as higher temperatures increase lattice anharmonicity, which in turn enhances higher-order phonon scattering processes such as four-phonon interactions. To quantify this behavior, Figs. \ref{fig:kappa_scatter}c and \ref{fig:kappa_scatter}e compare the total phonon scattering rates with and without four-phonon scattering at 77 K and 900 K, respectively. At 77 K, the scattering rates from both cases nearly overlap; in contrast, at 900 K, the inclusion of four-phonon scattering leads to a clear increase in the total scattering rate, particularly for high-frequency modes. To dive into the phonon scattering rates, Figure \ref{fig:kappa_scatter}c and \ref{fig:kappa_scatter}e compare the phonon scattering rate with and without four phonon scattering for 77 K and 900 K, respectively. It can be seen that the difference in scattering rate is obvious for 900 K while at 77 K, the values almost overlap. This temperature-dependent behavior can be theoretically explained by the scaling of phonon scattering rates with temperature. For phonons of a given frequency, the three-phonon scattering rate scales linearly with temperature ($\tau^{-1}_\mathrm{3ph} \propto T $),  while the four-phonon scattering rate increases quadratically ($\tau^{-1}_\mathrm{4ph} \propto T^2 $) \cite{Feng2017}. As a result, four-phonon scattering must be included to accurately capture the thermal resistance for high temperature applications.

Due to the metallic nature of UN, electrons play a significant role in thermal transport. In addition to phonon–phonon interactions, electron–phonon scattering serves as an important mechanism that limits $\kappa_\mathrm{L}$. To quantify this effect, we calculated electron–phonon coupling, which was then incorporated into the phonon scattering rates for evaluating  $\kappa_\mathrm{L}$. Figure \ref{fig:kappa_scatter}b compares the lattice thermal conductivity calculated with and without electron–phonon scattering, both on top of the already-included four-phonon scattering. A substantial difference is observed at low temperatures: electron–phonon interactions significantly suppress $\kappa_\mathrm{L}$ (e.g., reducing it by 80\% at 77 K). However, as temperature increases, the influence of electron–phonon scattering rapidly diminishes ( $>\sim$600 K). This temperature-dependent behavior is consistent with theoretical expectations. At low temperatures, the phonon population is limited to low-frequency modes, which couple more effectively with electrons. As temperature increases, phonon–phonon scattering (especially four-phonon processes) becomes the dominant resistive mechanism, overwhelming the influence of electron–phonon interactions. As a validation, we plot the total phonon scattering rates at 77 K and 900 K in Figs. \ref{fig:kappa_scatter}d and \ref{fig:kappa_scatter}f, respectively. At 77 K, the inclusion of electron–phonon scattering substantially increases the total phonon scattering rate across the whole frequency range. At 900 K, the scattering rates with and without electron–phonon interactions are nearly identical, confirming that phonon–phonon processes dominate in this regime.

\begin{figure}[htbp]
    \centering
    \includegraphics[width=0.88\linewidth]{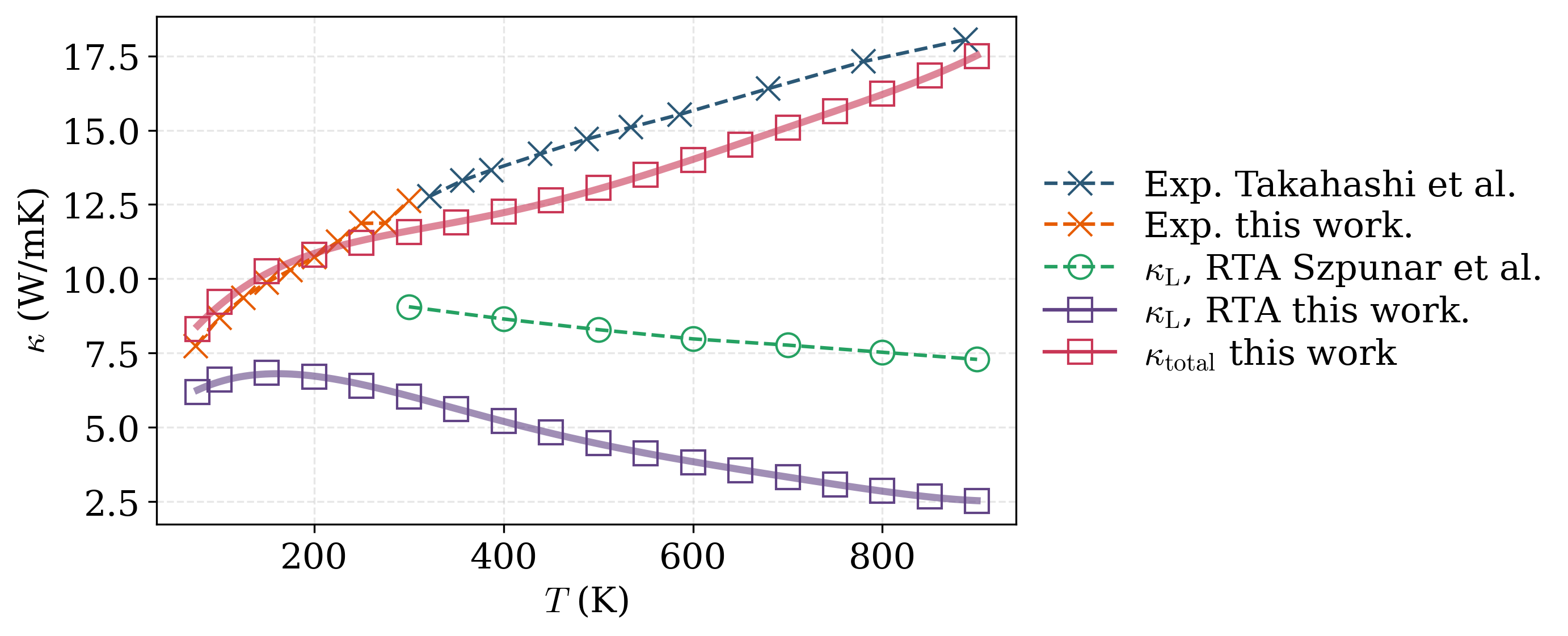}
    \caption{Predicted total thermal conductivity ($\kappa_\mathrm{total}$) compared with experimental data measured in this work (below 300 K) and taken from Takahashi et al. \cite{Takahashi1971} The $\kappa_\mathrm{L}$ from this work, which considers three-phonon, four-phonon, and electron-phonon scattering, is also compared with the results taken from Szpunar et al. \cite{Szpunar2020}, which consider three-phonon scattering. Calculations are based on the primitive 5$\times$5$\times$5 UN supercell.} 
    \label{fig:kappa_summary}
\end{figure}

Given the metallic nature of UN, electrons play a major role in thermal transport, particularly at high temperatures. Here, the total thermal conductivity of UN is evaluated by combining $\kappa_\mathrm{L}$, which accounts for three-phonon, four-phonon, and electron-phonon scattering, with $\kappa_\mathrm{e}$ evaluated from Eq \ref{eq:kappa_e} and \ref{eq:rho}. As shown in Fig. \ref{fig:kappa_summary}, the resulting $\kappa_\mathrm{total}$ generally shows good agreement with experimental data over a broad temperature range from 77 K to 900 K. The agreement is particularly strong below 300 K. Between 300 K and 800 K, $\kappa_\mathrm{total}$ is slightly lower than the experimental values, but the discrepancy diminishes again at temperatures above 800 K. A key contribution of this work is the inclusion of new thermal conductivity measurements using the SDTR technique in the range of 77 K to 300 K. These measurements address a critical gap in the literature, as existing experimental data, such as those by Takahashi et al. \cite{Takahashi1971}, lack low-temperature measurements. Our new addition fills in the data gap and also provides validation for the computational predictions in the low-temperature regime, where we found lattice thermal conductivity dominates and electron–phonon interactions play a significant role. This comparison with experimental data across the whole temperature range strengthens the reliability of the developed MTP for thermal transport applications.

For further comparison, Fig. \ref{fig:kappa_summary} also includes lattice thermal conductivity data from Szpunar et al. \cite{Szpunar2020}, obtained using RTA based on DFT with three-phonon scattering. While their study offered early insights into phonon transport in UN, their results show some inconsistencies across temperature, likely due to the omission of higher-order scattering processes. The improved agreement observed in our results highlights the importance of including both four-phonon and electron–phonon interactions for a more complete description.

Our findings also clarify the evolving role of phonons and electrons with temperature. Below $\sim$300 K, $\kappa_\mathrm{L}$ is the primary contributor to heat transport, but it is significantly reduced by electron–phonon scattering. As temperature increases, phonon–phonon scattering, especially four-phonon processes, further lowers $\kappa_\mathrm{L}$, while $\kappa_\mathrm{e}$ steadily increases. Above $\sim$800 K, electronic transport dominates, consistent with UN's metallic nature.

\section{Conclusion}

In this work, a MLIP based on the MTP framework was developed for predicting the thermal properties of UN. The MTP was trained using DFT-generated data and initially validated through the prediction of physical properties, including the equation of state, lattice constant, elastic constants, phonon dispersion, and defect formation energies. These results were benchmarked against DFT calculations, available experimental data, and predictions from other interatomic potentials. Overall, the developed MTP demonstrates balanced performance across a wide range of properties, avoiding the specific biases observed in other potential models.

The validated MTP was then employed to predict the thermal properties of UN, including melting point, thermal expansion, specific heat, and thermal conductivity. The results show close agreement with available experimental data. Notably, by incorporating new thermal conductivity measurements on single-crystalline UN, the study identifies the significant roles of four-phonon and electron-phonon scattering processes in accurately capturing thermal conductivity over a broad temperature range. Overall, the developed MTP enables MD simulations and LD calculations with quantum-level accuracy at much lower computational cost. It therefore offers a robust tool for future studies of defect behavior and high-temperature energy transport in UN.


\section*{Supplementary Material}
See the supplementary material containing the convergence tests for VASP setup and cell size in this work.

\section*{Acknowledgments}
This work is supported by the Center for Thermal Energy Transport under Irradiation, an Energy Frontier Research Center funded by the U.S. Department of Energy, Office of Science, Office of Basic Energy Sciences.

\section*{DATA AVAILABILITY}
The data that support the findings of this study are available from the corresponding author upon reasonable request.

\bibliographystyle{unsrt}
\bibliography{references}

@book{Srivastava2022,
   author = {Gyaneshwar P. Srivastava},
   city = {Boca Raton},
   doi = {10.1201/9781003141273},
   isbn = {9781003141273},
   month = {8},
   publisher = {CRC Press},
   title = {The Physics of Phonons},
   url = {https://www.taylorfrancis.com/books/9781003141273},
   year = {2022}
}

@article{McGaughey2004,


   author = {A. J. H. McGaughey and M. Kaviany},
   doi = {10.1103/PhysRevB.69.094303},
   issn = {1098-0121},
   issue = {9},
   journal = {Physical Review B},
   month = {3},
   pages = {094303},
   title = {Quantitative validation of the {B}oltzmann transport equation phonon thermal conductivity model under the single-mode relaxation time approximation},
   volume = {69},
   url = {https://link.aps.org/doi/10.1103/PhysRevB.69.094303},
   year = {2004}
}

@article{Ross1988,
   author = {Steven B. Ross and Mohamed S. El-Genk and R. Bruce Matthews},
   doi = {10.1016/0022-3115(88)90026-8},
   issn = {0022-3115},
   issue = {3},
   journal = {Journal of Nuclear Materials},
   month = {2},
   pages = {318-326},
   publisher = {North-Holland},
   title = {Thermal conductivity correlation for uranium nitride fuel between 10 and 1923 {K}},
   volume = {151},
   year = {1988},
}

@article{Hayes1990cp,
   author = {S. L. Hayes and J. K. Thomas and K. L. Peddicord},
   doi = {10.1016/0022-3115(90)90377-Y},
   issn = {00223115},
   issue = {2-3},
   journal = {Journal of Nuclear Materials},
   title = {Material property correlations for uranium mononitride. IV. Thermodynamic properties},
   volume = {171},
   year = {1990},
}

@article{Szpunar2020,
   author = {Barbara Szpunar and Jayangani I. Ranasinghe and Linu Malakkal and Jerzy A. Szpunar},
   doi = {10.1016/j.jpcs.2020.109636},
   issn = {00223697},
   journal = {Journal of Physics and Chemistry of Solids},
   month = {11},
   pages = {109636},
   title = {First principles investigation of thermal transport of uranium mononitride},
   volume = {146},
   year = {2020},
}

@article{AbdulHameed2024,
   author = {Mohamed AbdulHameed and Benjamin Beeler and Conor O.T. Galvin and Michael W.D. Cooper},
   doi = {10.1016/J.JNUCMAT.2024.155247},
   issn = {0022-3115},
   journal = {Journal of Nuclear Materials},
   month = {6},
   pages = {155247},
   publisher = {North-Holland},
   title = {Assessment of uranium nitride interatomic potentials},
   year = {2024},
}

@article{Tseplyaev2016,
   author = {V. I. Tseplyaev and S. V. Starikov},
   doi = {10.1016/J.JNUCMAT.2016.07.048},
   issn = {0022-3115},
   journal = {Journal of Nuclear Materials},
   month = {11},
   pages = {7-14},
   publisher = {North-Holland},
   title = {The atomistic simulation of pressure-induced phase transition in uranium mononitride},
   volume = {480},
   year = {2016},
}

@article{Kocevski2022,
   author = {Vancho Kocevski and Michael W.D. Cooper and Antoine J. Claisse and David A. Andersson},
   doi = {10.1016/j.jnucmat.2022.153553},
   issn = {00223115},
   journal = {Journal of Nuclear Materials},
   month = {4},
   pages = {153553},
   title = {Development and application of a uranium mononitride ({UN}) potential: Thermomechanical properties and {Xe} diffusion},
   volume = {562},
   year = {2022},
}

@article{Podryabinkin2023,
   author = {Evgeny Podryabinkin and Kamil Garifullin and Alexander Shapeev and Ivan Novikov},
   doi = {10.1063/5.0155887},
   issn = {0021-9606},
   issue = {8},
   journal = {The Journal of Chemical Physics},
   month = {8},
   title = {{MLIP-3}: Active learning on atomic environments with moment tensor potentials},
   volume = {159},
   url = {https://pubs.aip.org/jcp/article/159/8/084112/2908187/MLIP-3-Active-learning-on-atomic-environments-with},
   year = {2023},
}

@article{Jackman1986,
   author = {J. A. Jackman and T. M. Holden and W. J. L. Buyers and P. de V. DuPlessis and O. Vogt and J. Genossar},
   doi = {10.1103/PhysRevB.33.7144},
   issn = {0163-1829},
   issue = {10},
   journal = {Physical Review B},
   month = {5},
   pages = {7144-7153},
   title = {Systematic study of the lattice dynamics of the uranium rocksalt-structure compounds},
   volume = {33},
   url = {https://link.aps.org/doi/10.1103/PhysRevB.33.7144},
   year = {1986},
}

@article{Salleh1986,
   author = {M. D. Salleh and J. E. MacDonald and G. A. Saunders and P. De V. Du Plessis},
   doi = {10.1007/BF01114310},
   issn = {0022-2461},
   issue = {7},
   journal = {Journal of Materials Science},
   month = {7},
   pages = {2577-2580},
   title = {Hydrostatic pressure dependences of elastic constants and vibrational anharmonicity of uranium nitride},
   volume = {21},
   url = {http://link.springer.com/10.1007/BF01114310},
   year = {1986},
}

@article{Hayes1990gp,
   author = {S.L. Hayes and J.K. Thomas and K.L. Peddicord},
   doi = {10.1016/0022-3115(90)90375-W},
   issn = {00223115},
   issue = {2-3},
   journal = {Journal of Nuclear Materials},
   month = {5},
   pages = {271-288},
   title = {Material property correlations for uranium mononitride: {II}. Mechanical properties},
   volume = {171},
   url = {https://linkinghub.elsevier.com/retrieve/pii/002231159090375W},
   year = {1990},
}

@article{CARVAJALNUNEZ20141,
title = {Melting point determination of uranium nitride and uranium plutonium nitride: A laser heating study},
journal = {Journal of Nuclear Materials},
volume = {449},
number = {1},
pages = {1-8},
year = {2014},
issn = {0022-3115},
doi = {https://doi.org/10.1016/j.jnucmat.2014.02.021},
url = {https://www.sciencedirect.com/science/article/pii/S002231151400083X},
author = {U. {Carvajal Nunez} and D. Prieur and R. Bohler and D. Manara}
}

@article{Takahashi1971,
   author = {Y. Takahashi and M. Murabayashi and Y. Akimoto and T. Mukaibo},
   doi = {10.1016/0022-3115(71)90059-6},
   issn = {00223115},
   issue = {3},
   journal = {Journal of Nuclear Materials},
   month = {3},
   pages = {303-308},
   title = {Uranium mononitride: Heat capacity and thermal conductivity from 298 to 1000 {K}},
   volume = {38},
   url = {https://linkinghub.elsevier.com/retrieve/pii/0022311571900596},
   year = {1971},
}

@article{Shapeev2016,
   author = {Alexander V. Shapeev},
   doi = {10.1137/15M1054183},
   issn = {1540-3459},
   issue = {3},
   journal = {Multiscale Modeling and Simulation},
   month = {1},
   pages = {1153-1173},
   title = {Moment Tensor Potentials: A Class of Systematically Improvable Interatomic Potentials},
   volume = {14},
   url = {http://epubs.siam.org/doi/10.1137/15M1054183},
   year = {2016},
}

@techReport{Miller2024,
   author = {Zachary Miller and Alex Levinsky and Galen Craven and Vedant Mehta and Massimiliano Fratoni and Joshua White and Anders Andersson and Maria Kosmidou and Adrien Terricabras},
   city = {Los Alamos, NM (United States)},
   doi = {10.2172/2440180},
   institution = {Los Alamos National Laboratory (LANL)},
   month = {9},
   title = {Uranium Mononitride ({UN}) Handbook},
   url = {https://www.osti.gov/servlets/purl/2440180/},
   year = {2024}
}

@article{LIU2023154215,
title = {Thermal expansion and steam oxidation of uranium mononitride analysed via in situ neutron diffraction},
journal = {Journal of Nuclear Materials},
volume = {575},
pages = {154215},
year = {2023},
issn = {0022-3115},
doi = {https://doi.org/10.1016/j.jnucmat.2022.154215},
url = {https://www.sciencedirect.com/science/article/pii/S0022311522006948},
author = {Jiatu Liu and Claudia Gasparrini and Joshua T. White and Kyle Johnson and Denise Adorno Lopes and Vanessa K. Peterson and Andrew Studer and Grant J. Griffiths and Gregory R. Lumpkin and Mark R. Wenman and Patrick A. Burr and Elizabeth S. Sooby and Edward G. Obbard}
}

@article{Kocevski2022_spin,
   author = {Vancho Kocevski and Daniel A. Rehn and Michael W.D. Cooper and David A. Andersson},
   doi = {10.1016/j.jnucmat.2021.153401},
   issn = {00223115},
   journal = {Journal of Nuclear Materials},
   month = {2},
   pages = {153401},
   title = {First-principles investigation of uranium mononitride ({UN}): Effect of magnetic ordering, spin-orbit interactions and exchange correlation functional},
   volume = {559},
   url = {https://linkinghub.elsevier.com/retrieve/pii/S0022311521006218},
   year = {2022},
}

@article{LAMMPS,
  author = "A. P. Thompson and H. M. Aktulga and R. Berger and 
     D. S. Bolintineanu and W. M. Brown and P. S. Crozier and
     P. J. in 't Veld and A. Kohlmeyer and S. G. Moore and T. D. Nguyen and
     R. Shan and M. J. Stevens and J. Tranchida and C. Trott and S. J. Plimpton",
  title = "{LAMMPS} - a flexible simulation tool for
     particle-based materials modeling at the 
     atomic, meso, and continuum scales",
  journal = "Comp. Phys. Comm.",
  volume =  "271",
  pages =   "108171",
  year =    "2022",
  doi = "10.1016/j.cpc.2021.108171"
}

@article{PODRYABINKIN2017171,
title = {Active learning of linearly parametrized interatomic potentials},
journal = {Computational Materials Science},
volume = {140},
pages = {171-180},
year = {2017},
issn = {0927-0256},
doi = {https://doi.org/10.1016/j.commatsci.2017.08.031},
url = {https://www.sciencedirect.com/science/article/pii/S0927025617304536},
author = {Evgeny V. Podryabinkin and Alexander V. Shapeev}
}

@misc{novikov_interface_lammps_mlip_3,
  author       = {Novikov, Ivan},
  title        = {{LAMMPS-MLIP-3} interface},
  howpublished = {\url{https://gitlab.com/ivannovikov/interface-lammps-mlip-3}}
}

@article{Perdew1996,
   author = {John P. Perdew and Kieron Burke and Matthias Ernzerhof},
   doi = {10.1103/PhysRevLett.77.3865},
   issn = {0031-9007},
   issue = {18},
   journal = {Physical Review Letters},
   month = {10},
   pages = {3865-3868},
   title = {Generalized Gradient Approximation Made Simple},
   volume = {77},
   year = {1996},
}

@article{vasp1Kresse1996,
   author = {G. Kresse and J. Furthmüller},
   doi = {10.1103/PhysRevB.54.11169},
   issn = {0163-1829},
   issue = {16},
   journal = {Physical Review B},
   month = {10},
   pages = {11169-11186},
   title = {Efficient iterative schemes for ab initio total-energy calculations using a plane-wave basis set},
   volume = {54},
   year = {1996},
}

@article{vasp2Kresse1996,
   author = {G. Kresse and J. Furthmüller},
   doi = {10.1016/0927-0256(96)00008-0},
   issn = {09270256},
   issue = {1},
   journal = {Computational Materials Science},
   month = {7},
   pages = {15-50},
   title = {Efficiency of ab-initio total energy calculations for metals and semiconductors using a plane-wave basis set},
   volume = {6},
   year = {1996},
}

@article{phonopy-phono3py-JPCM,
  author  = {Togo, Atsushi and Chaput, Laurent and Tadano, Terumasa and Tanaka, Isao},
  title   = {Implementation strategies in phonopy and phono3py},
  journal = {J. Phys. Condens. Matter},
  volume  = {35},
  number  = {35},
  pages   = {353001},
  year    = {2023},
  doi     = {10.1088/1361-648X/acd831}
}

@article{phonopy-phono3py-JPSJ,
  author  = {Togo, Atsushi},
  title   = {First-principles Phonon Calculations with {Phonopy} and {Phono3py}},
  journal = {J. Phys. Soc. Jpn.},
  volume  = {92},
  number  = {1},
  pages   = {012001},
  year    = {2023},
  doi     = {10.7566/JPSJ.92.012001}
}

@article{Zuo_2020, title={Performance and cost assessment of Machine learning interatomic potentials}, volume={124}, url={https://pubs.acs.org/doi/10.1021/acs.jpca.9b08723}, DOI={10.1021/acs.jpca.9b08723}, number={4}, journal={The Journal of Physical Chemistry A}, author={Zuo, Yunxing and Chen, Chi and Li, Xiangguo and Deng, Zhi and Chen, Yiming and Behler, Jörg and Csányi, Gábor and Shapeev, Alexander V. and Thompson, Aidan P. and Wood, Mitchell A. and Ong, Shyue Ping}, year={2020}, month=jan, pages={731–745} }

@article{Ekberg2018, title={Nitride fuel for {Gen} {IV} nuclear power systems}, volume={318}, url={https://link.springer.com/article/10.1007/s10967-018-6316-0}, DOI={10.1007/s10967-018-6316-0}, number={3}, journal={Journal of Radioanalytical and Nuclear Chemistry}, author={Ekberg, Christian and Costa, Diogo Ribeiro and Hedberg, Marcus and Jolkkonen, Mikael}, year={2018}, month=nov, pages={1713–1725} }

@article{Hayes1990,
   author = {S.L. Hayes and J.K. Thomas and K.L. Peddicord},
   doi = {10.1016/0022-3115(90)90374-V},
   issn = {00223115},
   issue = {2-3},
   journal = {Journal of Nuclear Materials},
   month = {5},
   pages = {262-270},
   title = {Material property correlations for uranium mononitride: I. Physical properties},
   volume = {171},
   url = {https://linkinghub.elsevier.com/retrieve/pii/002231159090374V},
   year = {1990}
}

@article{Hayes1990tc,
   author = {S.L. Hayes and J.K. Thomas and K.L. Peddicord},
   doi = {10.1016/0022-3115(90)90376-X},
   issn = {00223115},
   issue = {2-3},
   journal = {Journal of Nuclear Materials},
   month = {5},
   pages = {289-299},
   title = {Material property correlations for uranium mononitride: {III}. Transport properties},
   volume = {171},
   url = {https://linkinghub.elsevier.com/retrieve/pii/002231159090376X},
   year = {1990},
}

@article{Han2022,
   author = {Zherui Han and Xiaolong Yang and Wu Li and Tianli Feng and Xiulin Ruan},
   doi = {10.1016/j.cpc.2021.108179},
   issn = {00104655},
   journal = {Computer Physics Communications},
   month = {1},
   pages = {108179},
   title = {{FourPhonon}: An extension module to ShengBTE for computing four-phonon scattering rates and thermal conductivity},
   volume = {270},
   url = {https://linkinghub.elsevier.com/retrieve/pii/S0010465521002915},
   year = {2022}
}

@misc{guo2023,
      title={Sampling-accelerated First-principles Prediction of Phonon Scattering Rates for Converged Thermal Conductivity and Radiative Properties}, 
      author={Ziqi Guo and Zherui Han and Dudong Feng and Guang Lin and Xiulin Ruan},
      year={2023},
      eprint={2311.12935},
      archivePrefix={arXiv},
      primaryClass={cond-mat.mtrl-sci},
      url={https://arxiv.org/abs/2311.12935}, 
}

@article{Kurosaki2000,
   author = {Ken Kurosaki and Kimihiko Yano and Kazuhiro Yamada and Masayoshi Uno and Shinsuke Yamanaka},
   doi = {10.1016/S0925-8388(00)01127-0},
   issn = {09258388},
   issue = {2},
   journal = {Journal of Alloys and Compounds},
   month = {10},
   pages = {305-310},
   title = {A molecular dynamics study of the thermal conductivity of uranium mononitride},
   volume = {311},
   year = {2000}
}

@article{Kurosaki2000sp,
   author = {Ken Kurosaki and Kimihiko Yano and Kazuhiro Yamada and Masayoshi Uno and Shinsuke Yamanaka},
   doi = {10.1016/S0925-8388(99)00561-7},
   issn = {09258388},
   issue = {1-2},
   journal = {Journal of Alloys and Compounds},
   month = {2},
   pages = {1-4},
   title = {A molecular dynamics study of the heat capacity of uranium mononitride},
   volume = {297},
   url = {https://linkinghub.elsevier.com/retrieve/pii/S0925838899005617},
   year = {2000}
}

@article{Chen2010,
   author = {P. H. Chen and X. L. Wang and X. C. Lai and G. Li and B. Y. Ao and Y. Long},
   doi = {10.1016/J.JNUCMAT.2010.06.017},
   issn = {0022-3115},
   issue = {1},
   journal = {Journal of Nuclear Materials},
   month = {9},
   pages = {6-8},
   publisher = {North-Holland},
   title = {Ab initio interionic potentials for {UN} by multiple lattice inversion},
   volume = {404},
   year = {2010}
}

@article{Kuksin2016,
   author = {A.Yu. Kuksin and S.V. Starikov and D.E. Smirnova and V.I. Tseplyaev},
   doi = {10.1016/j.jallcom.2015.10.223},
   issn = {09258388},
   journal = {Journal of Alloys and Compounds},
   month = {2},
   pages = {385-394},
   publisher = {Elsevier},
   title = {The diffusion of point defects in uranium mononitride: Combination of {DFT} and atomistic simulation with novel potential},
   volume = {658},
   url = {https://linkinghub.elsevier.com/retrieve/pii/S0925838815314791},
   year = {2016}
}

@article{Mei2013,
   author = {Zhi-Gang Mei and Marius Stan and Benjamin Pichler},
   doi = {10.1016/j.jnucmat.2013.04.058},
   issn = {00223115},
   issue = {1-3},
   journal = {Journal of Nuclear Materials},
   month = {9},
   pages = {63-69},
   title = {First-principles study of structural, elastic, electronic, vibrational and thermodynamic properties of {UN}},
   volume = {440},
   year = {2013}
}

@article{Gryaznov2012,
   author = {Denis Gryaznov and Eugene Heifets and Eugene Kotomin},
   doi = {10.1039/c2cp40297a},
   issn = {1463-9076},
   issue = {13},
   journal = {Physical Chemistry Chemical Physics},
   pages = {4482},
   title = {The first-principles treatment of the electron-correlation and spin–orbital effects in uranium mononitride nuclear fuels},
   volume = {14},
   url = {https://xlink.rsc.org/?DOI=c2cp40297a},
   year = {2012}
}

@article{Evarestov2008,
   author = {R. A. Evarestov and M. V. Losev and A. I. Panin and N. S. Mosyagin and A. V. Titov},
   doi = {10.1002/pssb.200743247},
   issn = {0370-1972},
   issue = {1},
   journal = {physica status solidi (b)},
   month = {1},
   pages = {114-122},
   title = {Electronic structure of crystalline uranium nitride: {LCAO DFT} calculations},
   volume = {245},
   year = {2008}
}

@article{Jones2023,
   author = {Suzanne Jones and Colin Boxall and Chris Maher and Robin Taylor},
   doi = {10.1016/J.PNUCENE.2023.104917},
   issn = {0149-1970},
   journal = {Progress in Nuclear Energy},
   month = {11},
   pages = {104917},
   publisher = {Pergamon},
   title = {A review of the reprocessability of uranium nitride based fuels},
   volume = {165},
   year = {2023}
}

@article{Youinou2014,
   author = {Gilles J. Youinou and R. Sonat Sen},
   doi = {10.13182/NT14-22},
   issn = {19437471},
   issue = {2},
   journal = {Nuclear Technology},
   month = {11},
   pages = {123-138},
   publisher = {Taylor & Francis},
   title = {Impact of Accident-Tolerant Fuels and Claddings on the Overall Fuel Cycle: A Preliminary Systems Analysis},
   volume = {188},
   url = {https://www.tandfonline.com/doi/abs/10.13182/NT14-22},
   year = {2014}
}

@article{Gueddim2010,
   author = {A. Gueddim and N. Bouarissa and A. Villesuzanne},
   doi = {10.1016/J.COMMATSCI.2010.02.010},
   issn = {0927-0256},
   issue = {3},
   journal = {Computational Materials Science},
   month = {5},
   pages = {490-494},
   publisher = {Elsevier},
   title = {Pressure dependence of elastic constants and related parameters for rocksalt {MgO}},
   volume = {48},
   year = {2010}
}

@article{Behler2007,
   author = {Jörg Behler and Michele Parrinello},
   doi = {10.1103/PHYSREVLETT.98.146401/FIGURES/4/MEDIUM},
   issn = {00319007},
   issue = {14},
   journal = {Physical Review Letters},
   month = {4},
   pages = {146401},
   pmid = {17501293},
   publisher = {American Physical Society},
   title = {Generalized neural-network representation of high-dimensional potential-energy surfaces},
   volume = {98},
   url = {https://journals.aps.org/prl/abstract/10.1103/PhysRevLett.98.146401},
   year = {2007}
}

@article{Behler2011,
   author = {Jörg Behler},
   doi = {10.1039/c1cp21668f},
   issn = {1463-9076},
   issue = {40},
   journal = {Physical Chemistry Chemical Physics},
   month = {10},
   pages = {17930},
   pmid = {21915403},
   publisher = {Royal Society of Chemistry},
   title = {Neural network potential-energy surfaces in chemistry: a tool for large-scale simulations},
   volume = {13},
   url = {https://xlink.rsc.org/?DOI=c1cp21668f},
   year = {2011}
}

@article{Bartk2013,
   author = {Albert P. Bartók and Risi Kondor and Gábor Csányi},
   doi = {10.1103/PhysRevB.87.184115},
   issn = {1098-0121},
   issue = {18},
   journal = {Physical Review B},
   month = {5},
   pages = {184115},
   publisher = {American Physical Society},
   title = {On representing chemical environments},
   volume = {87},
   url = {https://link.aps.org/doi/10.1103/PhysRevB.87.184115},
   year = {2013}
}

@article{Bartk2010,
   author = {Albert P. Bartók and Mike C. Payne and Risi Kondor and Gábor Csányi},
   doi = {10.1103/PhysRevLett.104.136403},
   issn = {0031-9007},
   issue = {13},
   journal = {Physical Review Letters},
   month = {4},
   pages = {136403},
   pmid = {20481899},
   publisher = {American Physical Society},
   title = {Gaussian Approximation Potentials: The Accuracy of Quantum Mechanics, without the Electrons},
   volume = {104},
   url = {https://link.aps.org/doi/10.1103/PhysRevLett.104.136403},
   year = {2010}
}

@article{Thompson2015,
   author = {A. P. Thompson and L. P. Swiler and C. R. Trott and S. M. Foiles and G. J. Tucker},
   doi = {10.1016/J.JCP.2014.12.018},
   issn = {0021-9991},
   journal = {Journal of Computational Physics},
   month = {3},
   pages = {316-330},
   publisher = {Academic Press},
   title = {Spectral neighbor analysis method for automated generation of quantum-accurate interatomic potentials},
   volume = {285},
   year = {2015}
}

@article{Li2018,
   author = {Xiang Guo Li and Chongze Hu and Chi Chen and Zhi Deng and Jian Luo and Shyue Ping Ong},
   doi = {10.1103/PHYSREVB.98.094104/SI.PDF},
   issn = {24699969},
   issue = {9},
   journal = {Physical Review B},
   month = {9},
   pages = {094104},
   publisher = {American Physical Society},
   title = {Quantum-accurate spectral neighbor analysis potential models for {Ni-Mo} binary alloys and fcc metals},
   volume = {98},
   url = {https://journals.aps.org/prb/abstract/10.1103/PhysRevB.98.094104},
   year = {2018}
}

@article{Lan2013,
   author = {Jian-Hui Lan and Zi-Chen Zhao and Qiong Wu and Yu-Liang Zhao and Zhi-Fang Chai and Wei-Qun Shi},
   doi = {10.1063/1.4846217},
   issn = {0021-8979},
   issue = {22},
   journal = {Journal of Applied Physics},
   month = {12},
   pages = {223516},
   publisher = {AIP Publishing},
   title = {First-principles {DFT+U} modeling of defect behaviors in anti-ferromagnetic uranium mononitride},
   volume = {114},
   url = {https://pubs.aip.org/jap/article/114/22/223516/396550/First-principles-DFT-U-modeling-of-defect},
   year = {2013}
}

@inbook{Mishin2005,
   author = {Y. Mishin},
   city = {Dordrecht},
   doi = {10.1007/978-1-4020-3286-8_23},
   isbn = {978-1-4020-3286-8},
   booktitle = {Handbook of Materials Modeling},
   pages = {459-478},
   publisher = {Springer Netherlands},
   title = {Interatomic Potentials for Metals},
   url = {http://link.springer.com/10.1007/978-1-4020-3286-8_23},
   year = {2005}
}

@article{Anderson1963,
   author = {Orson L. Anderson},
   doi = {10.1016/0022-3697(63)90067-2},
   issn = {0022-3697},
   issue = {7},
   journal = {Journal of Physics and Chemistry of Solids},
   month = {7},
   pages = {909-917},
   publisher = {Pergamon},
   title = {A simplified method for calculating the Debye temperature from elastic constants},
   volume = {24},
   year = {1963}
}

@article{Hill1952,
   author = {R Hill},
   doi = {10.1088/0370-1298/65/5/307},
   issn = {0370-1298},
   issue = {5},
   journal = {Proceedings of the Physical Society. Section A},
   month = {5},
   pages = {349-354},
   title = {The Elastic Behaviour of a Crystalline Aggregate},
   volume = {65},
   year = {1952}
}

@article{Bowler2012,
   author = {D R Bowler and T Miyazaki},
   doi = {10.1088/0034-4885/75/3/036503},
   issn = {0034-4885},
   issue = {3},
   journal = {Reports on Progress in Physics},
   month = {3},
   pages = {036503},
   title = {{$\mathcal{O}$(N) methods in electronic structure calculations}},
   volume = {75},
   url = {https://iopscience.iop.org/article/10.1088/0034-4885/75/3/036503},
   year = {2012}
}

@article{Bock2024,
   author = {F. Bock and F. Tasnádi and I. A. Abrikosov},
   doi = {10.1116/6.0003260},
   issn = {0734-2101},
   issue = {1},
   journal = {Journal of Vacuum Science \& Technology A},
   month = {1},
   title = {Active learning with moment tensor potentials to predict material properties: {Ti0.5Al0.5N} at elevated temperature},
   volume = {42},
   url = {https://pubs.aip.org/jva/article/42/1/013412/2932400/Active-learning-with-moment-tensor-potentials-to},
   year = {2024}
}

@article{Franz1853,
   author = {R. Franz and G. Wiedemann},
   doi = {10.1002/andp.18531650802},
   issn = {0003-3804},
   issue = {8},
   journal = {Annalen der Physik},
   month = {1},
   pages = {497-531},
   title = {Ueber die Wärme‐Leitungsfähigkeit der Metalle},
   volume = {165},
   year = {1853}
}

@Article{Lee2023,
   Title   = {Electron--phonon physics from first principles using the {EPW} code},
   Author  = {H. Lee and S. Ponc\'e and K. Bushick and S. Hajinazar and J. Lafuente-Bartolome and J. Leveillee and C. Lian and J. Lihm and F. Macheda and H. Mori and H. Paudyal and W.H. Sio and S. Tiwari and M. Zacharias and X. Zhang and N. Bonini, Nicola and E. Kioupakis},
   Journal = {npj Computational Materials},
   Year    = {2023},
   Volume  = {9},
   Pages   = {156},
   Doi     = {10.1038/s41524-023-01107-3}
 }

@Article{Ponce2016,
   Title   = {{EPW}: Electron–phonon coupling, transport and superconducting properties using maximally localized {Wannier} functions},
   Author  = {S. Ponc\'e and E.R. Margine and C. Verdi and F. Giustino},
   Journal = {Computer Physics Communications},
   Year    = {2016},
   Volume  = {209},
   Pages   = {116 - 133},
   Doi     = {https://doi.org/10.1016/j.cpc.2016.07.028}
 }

@article{mehl1994,
  title={First principles calculations of elastic properties of metals},
  author={Mehl, Michael J and Klein, Barry M and Papaconstantopoulos, Dimitri A and others},
  journal={Intermetallic compounds: principles and practice},
  volume={1},
  pages={195--210},
  year={1994},
  publisher={Citeseer}
}

@article{Feng2017,
   author = {Tianli Feng and Lucas Lindsay and Xiulin Ruan},
   doi = {10.1103/PhysRevB.96.161201},
   issn = {2469-9950},
   issue = {16},
   journal = {Physical Review B},
   month = {10},
   pages = {161201},
   title = {Four-phonon scattering significantly reduces intrinsic thermal conductivity of solids},
   volume = {96},
   year = {2017}
}

@article{Oetting1972,
   author = {Franklin L. Oetting and James M. Leitnaker},
   doi = {10.1016/0021-9614(72)90057-2},
   issn = {00219614},
   issue = {2},
   journal = {The Journal of Chemical Thermodynamics},
   month = {3},
   pages = {199-211},
   title = {The chemical thermodynamic properties of nuclear materials {I}. Uranium mononitride},
   volume = {4},
   year = {1972}
}

@article{Marples1970,
   author = {J.A.C. Marples},
   doi = {10.1016/0022-3697(70)90060-0},
   issn = {00223697},
   issue = {11},
   journal = {Journal of Physics and Chemistry of Solids},
   month = {11},
   pages = {2431-2439},
   title = {An {X-ray} study of the structures of {UN}, {UP}, {US} and {USe} at cryogenic temperatures},
   volume = {31},
   url = {https://linkinghub.elsevier.com/retrieve/pii/0022369770900600},
   year = {1970}
}

@article{murnaghan,
  title={The compressibility of media under extreme pressures},
  author={Murnaghan, Francis Dominic},
  journal={Proceedings of the National Academy of Sciences},
  volume={30},
  number={9},
  pages={244--247},
  year={1944}
}

@article{Esfarjani2008,
   author = {Keivan Esfarjani and Harold T. Stokes},
   doi = {10.1103/PhysRevB.77.144112},
   issn = {1098-0121},
   issue = {14},
   journal = {Physical Review B},
   month = {4},
   pages = {144112},
   title = {Method to extract anharmonic force constants from first principles calculations},
   volume = {77},
   url = {https://link.aps.org/doi/10.1103/PhysRevB.77.144112},
   year = {2008}
}

@article{Feng2018,
   author = {Tianli Feng and Xiaolong Yang and Xiulin Ruan},
   doi = {10.1063/1.5048799},
   issn = {0021-8979},
   issue = {14},
   journal = {Journal of Applied Physics},
   month = {10},
   title = {Phonon anharmonic frequency shift induced by four-phonon scattering calculated from first principles},
   volume = {124},
   url = {https://pubs.aip.org/jap/article/124/14/145101/155500/Phonon-anharmonic-frequency-shift-induced-by-four},
   year = {2018}
}

@article{Hurley2015,
   author = {David H. Hurley and Robert S. Schley and Marat Khafizov and Brycen L. Wendt},
   doi = {10.1063/1.4936213},
   issn = {0034-6748},
   issue = {12},
   journal = {Review of Scientific Instruments},
   month = {12},
   title = {Local measurement of thermal conductivity and diffusivity},
   volume = {86},
   url = {https://pubs.aip.org/rsi/article/86/12/123901/857179/Local-measurement-of-thermal-conductivity-and},
   year = {2015}
}

@article{Hua2012,
   author = {Zilong Hua and Heng Ban and Marat Khafizov and Robert Schley and Rory Kennedy and David H. Hurley},
   doi = {10.1063/1.4716474},
   issn = {0021-8979},
   issue = {10},
   journal = {Journal of Applied Physics},
   month = {5},
   title = {Spatially localized measurement of thermal conductivity using a hybrid photothermal technique},
   volume = {111},
   url = {https://pubs.aip.org/jap/article/111/10/103505/369344/Spatially-localized-measurement-of-thermal},
   year = {2012}
}

@misc{Yurgens2010,
  author       = {A. Yurgens},
  title        = {Temperature distribution in large Bi2212 mesas},
  year         = {2010},
  month        = may,
  note         = {arXiv:1005.2932},
  url          = {https://arxiv.org/pdf/1005.2932},
  archivePrefix= {arXiv},
  eprint       = {1005.2932}
}

@article{Deskins2022,
   author = {W. Ryan Deskins and Amey Khanolkar and Sanjoy Mazumder and Cody A. Dennett and Kaustubh Bawane and Zilong Hua and Joshua Ferrigno and Lingfeng He and J. Matthew Mann and Marat Khafizov and David H. Hurley and Anter El-Azab},
   doi = {10.1016/J.ACTAMAT.2022.118379},
   issn = {1359-6454},
   journal = {Acta Materialia},
   month = {12},
   pages = {118379},
   publisher = {Pergamon},
   title = {A combined theoretical-experimental investigation of thermal transport in low-dose irradiated thorium dioxide},
   volume = {241},
   url = {https://www.sciencedirect.com/science/article/pii/S1359645422007571?via%3Dihub},
   year = {2022}
}

@article{Dennett2020,
   author = {Cody A. Dennett and Zilong Hua and Amey Khanolkar and Tiankai Yao and Phyllis K. Morgan and Timothy A. Prusnick and Narayan Poudel and Aaron French and Krzysztof Gofryk and Lingfeng He and Lin Shao and Marat Khafizov and David B. Turner and J. Matthew Mann and David H. Hurley},
   doi = {10.1063/5.0025384/569830},
   issn = {2166532X},
   issue = {11},
   journal = {APL Materials},
   month = {11},
   pages = {111103},
   publisher = {American Institute of Physics Inc.},
   title = {The influence of lattice defects, recombination, and clustering on thermal transport in single crystal thorium dioxide},
   volume = {8},
   url = {/aip/apm/article/8/11/111103/569830/The-influence-of-lattice-defects-recombination-and},
   year = {2020}
}

@article{Hua2020,
   author = {Zilong Hua and Tiankai Yao and Amey Khanolkar and Xiaxin Ding and Krzysztof Gofryk and Lingfeng He and Michael Benson and David Hurley},
   doi = {10.1016/J.JNUCMAT.2020.152145},
   issn = {0022-3115},
   journal = {Journal of Nuclear Materials},
   month = {6},
   pages = {152145},
   publisher = {North-Holland},
   title = {Intragranular thermal transport in {U–50Zr}},
   volume = {534},
   url = {https://www.sciencedirect.com/science/article/pii/S0022311519316678?via%3Dihub},
   year = {2020}
}

@article{Lorenz1872,
   author = {L. Lorenz},
   doi = {10.1002/andp.18722231107},
   issn = {0003-3804},
   issue = {11},
   journal = {Annalen der Physik},
   month = {1},
   pages = {429-452},
   title = {Bestimmung der Wärmegrade in absolutem Maasse},
   volume = {223},
   year = {1872}
}

@article{alzate,
   author = {Lorena Alzate-Vargas and Kashi N Subedi and Nicholas Lubbers and Michael W D Cooper and Roxanne M Tutchton and Tammie Gibson and Richard A Messerly},
   doi = {10.1088/2632-2153/ae0242},
   issn = {2632-2153},
   issue = {3},
   journal = {Machine Learning: Science and Technology},
   month = {9},
   pages = {035064},
   title = {Toward machine learning interatomic potentials for modeling uranium mononitride},
   volume = {6},
   url = {https://iopscience.iop.org/article/10.1088/2632-2153/ae0242},
   year = {2025}
}

@article{TENNERY1971,
   author = {V. J. TENNERY and T. G. GODFREY and R. A. POTTER},
   doi = {10.1111/J.1151-2916.1971.TB12306.X;PAGE:STRING:ARTICLE/CHAPTER},
   issn = {15512916},
   issue = {7},
   journal = {Journal of the American Ceramic Society},
   month = {7},
   pages = {327-331},
   publisher = {John Wiley \& Sons, Ltd},
   title = {Sintering of {UN} as a Function of Temperature and {N$_2$} Pressure},
   volume = {54},
   url = {/doi/pdf/10.1111/j.1151-2916.1971.tb12306.x https://onlinelibrary.wiley.com/doi/abs/10.1111/j.1151-2916.1971.tb12306.x https://ceramics.onlinelibrary.wiley.com/doi/10.1111/j.1151-2916.1971.tb12306.x},
   year = {1971}
}

@article{Nos1984,
   author = {Shuichi Nosé},
   doi = {10.1063/1.447334},
   issn = {0021-9606},
   issue = {1},
   journal = {The Journal of Chemical Physics},
   month = {7},
   pages = {511-519},
   publisher = {AIP Publishing},
   title = {A unified formulation of the constant temperature molecular dynamics methods},
   volume = {81},
   url = {/aip/jcp/article/81/1/511/607222/A-unified-formulation-of-the-constant-temperature},
   year = {1984}
}

@article{Hoover1985,
   author = {William G. Hoover},
   doi = {10.1103/PhysRevA.31.1695},
   issn = {10502947},
   issue = {3},
   journal = {Physical Review A},
   month = {3},
   pages = {1695},
   pmid = {9895674},
   publisher = {American Physical Society},
   title = {Canonical dynamics: Equilibrium phase-space distributions},
   volume = {31},
   url = {https://journals.aps.org/pra/abstract/10.1103/PhysRevA.31.1695},
   year = {1985}
}

@article{Claisse2016,
   author = {Antoine Claisse and Marco Klipfel and Niclas Lindbom and Michel Freyss and Pär Olsson},
   doi = {10.1016/J.JNUCMAT.2016.06.007},
   issn = {0022-3115},
   journal = {Journal of Nuclear Materials},
   month = {9},
   pages = {119-124},
   publisher = {North-Holland},
   title = {{GGA+U} study of uranium mononitride: A comparison of the U-ramping and occupation matrix schemes and incorporation energies of fission products},
   volume = {478},
   url = {https://www.sciencedirect.com/science/article/pii/S0022311516302598#fig1},
   year = {2016}
}

@article{Samsel,
   author = {M. Samsel-Czekała and E. Talik and P. de V. Du Plessis and R. Troć and H. Misiorek and C. Sułkowski},
   doi = {10.1103/PhysRevB.76.144426},
   issn = {1098-0121},
   issue = {14},
   journal = {Physical Review B},
   month = {10},
   pages = {144426},
   title = {Electronic structure and magnetic and transport properties of single-crystalline {UN}},
   volume = {76},
   url = {https://link.aps.org/doi/10.1103/PhysRevB.76.144426},
   year = {2007}
}

@article{Mostofi2014,
   author = {Arash A. Mostofi and Jonathan R. Yates and Giovanni Pizzi and Young-Su Lee and Ivo Souza and David Vanderbilt and Nicola Marzari},
   doi = {10.1016/j.cpc.2014.05.003},
   issn = {00104655},
   issue = {8},
   journal = {Computer Physics Communications},
   month = {8},
   pages = {2309-2310},
   title = {An updated version of {Wannier}90: A tool for obtaining maximally-localised {Wannier} functions},
   volume = {185},
   year = {2014}
}

@article{Marzari2012,
   author = {Nicola Marzari and Arash A. Mostofi and Jonathan R. Yates and Ivo Souza and David Vanderbilt},
   doi = {10.1103/RevModPhys.84.1419},
   issn = {0034-6861},
   issue = {4},
   journal = {Reviews of Modern Physics},
   month = {10},
   pages = {1419-1475},
   title = {Maximally localized {Wannier} functions: Theory and applications},
   volume = {84},
   year = {2012}
}

@article{Togo2010,
   author = {Atsushi Togo and Laurent Chaput and Isao Tanaka and Gilles Hug},
   doi = {10.1103/PhysRevB.81.174301},
   issn = {1098-0121},
   issue = {17},
   journal = {Physical Review B},
   month = {5},
   pages = {174301},
   title = {First-principles phonon calculations of thermal expansion in {Ti$_3$SiC$_2$}, {Ti$_3$AlC$_2$}, and {Ti$_3$GeC$_2$}},
   volume = {81},
   year = {2010}
}

@article{vanGunsteren2024,
   author = {Wilfred F. van Gunsteren and Chris Oostenbrink},
   doi = {10.1021/acs.jcim.4c00823},
   issn = {1549-9596},
   issue = {16},
   journal = {Journal of Chemical Information and Modeling},
   month = {8},
   pages = {6281-6304},
   title = {Methods for Classical-Mechanical Molecular Simulation in Chemistry: Achievements, Limitations, Perspectives},
   volume = {64},
   url = {https://pubs.acs.org/doi/10.1021/acs.jcim.4c00823},
   year = {2024}
}

@article{Baroni2010,
   author = {S. Baroni and P. Giannozzi and E. Isaev},
   doi = {10.2138/rmg.2010.71.3},
   issn = {1529-6466},
   issue = {1},
   journal = {Reviews in Mineralogy and Geochemistry},
   month = {1},
   pages = {39-57},
   title = {Density-Functional Perturbation Theory for Quasi-Harmonic Calculations},
   volume = {71},
   year = {2010}
}

@article{Chevalier2000,
   author = {P.-Y. Chevalier and E. Fischer and B. Cheynet},
   doi = {10.1016/S0022-3115(00)00043-X},
   issn = {00223115},
   issue = {2},
   journal = {Journal of Nuclear Materials},
   month = {7},
   pages = {136-150},
   title = {Thermodynamic modelling of the {N–U} system},
   volume = {280},
   url = {https://linkinghub.elsevier.com/retrieve/pii/S002231150000043X},
   year = {2000}
}

@article{Govers2008,
   author = {K. Govers and S. Lemehov and M. Hou and M. Verwerft},
   doi = {10.1016/j.jnucmat.2008.01.023},
   issn = {00223115},
   issue = {1},
   journal = {Journal of Nuclear Materials},
   month = {5},
   pages = {66-77},
   publisher = {North-Holland},
   title = {Comparison of interatomic potentials for UO2},
   volume = {376},
   url = {https://linkinghub.elsevier.com/retrieve/pii/S0022311508000986},
   year = {2008}
}

@article{Green1954,
   author = {Melville S. Green},
   doi = {10.1063/1.1740082},
   issn = {0021-9606},
   issue = {3},
   journal = {The Journal of Chemical Physics},
   month = {3},
   pages = {398-413},
   title = {Markoff Random Processes and the Statistical Mechanics of Time‐Dependent Phenomena. II. Irreversible Processes in Fluids},
   volume = {22},
   url = {http://aip.scitation.org/doi/10.1063/1.1740082},
   year = {1954}
}

@article{Kubo1957,
   author = {Ryogo Kubo},
   doi = {10.1143/JPSJ.12.570},
   issn = {0031-9015},
   issue = {6},
   journal = {Journal of the Physical Society of Japan},
   month = {6},
   pages = {570-586},
   title = {Statistical-Mechanical Theory of Irreversible Processes. I. General Theory and Simple Applications to Magnetic and Conduction Problems},
   volume = {12},
   url = {https://journals.jps.jp/doi/10.1143/JPSJ.12.570},
   year = {1957}
}

\end{document}